\documentclass[trackchanges,twocolumn]{aastex7}

\newcommand{\T}{$T_{\rm eff}$}
\newcommand{\g}{log($g$)}

\providecommand{\gaia}{\textit{Gaia}}
\providecommand{\gdr}[1]{\textit{Gaia}~DR{#1}}

\usepackage{amsmath}	
\usepackage{amssymb}
\usepackage{listings}
\usepackage{gensymb}
\usepackage{natbib}
\usepackage[detect-all]{siunitx}

\usepackage{graphicx}

\usepackage{txfonts}

\defcitealias{Poggio:2025}{Paper I}
\defcitealias{Poggio:2021}{P21}

\begin{document}

\title{The flare and spiral structure of the Milky Way's disc as traced by young giant stars}

\author[0000-0003-3793-8505]{E. Poggio}
\affiliation{INAF - Osservatorio Astrofisico di Torino, via Osservatorio 20, 10025 Pino Torinese (TO), Italy}
\email[show]{eloisa.poggio@inaf.it}

\author[0000-0002-1777-5502]{R. Drimmel}
\affiliation{INAF - Osservatorio Astrofisico di Torino, via Osservatorio 20, 10025 Pino Torinese (TO), Italy}
\email[notshow]{eloisa.poggio@inaf.it}

\author[0000-0002-2604-4277]{S. Khanna}
\affiliation{INAF - Osservatorio Astrofisico di Torino, via Osservatorio 20, 10025 Pino Torinese (TO), Italy}
\email[notshow]{eloisa.poggio@inaf.it}

\author[0000-0001-8006-6365]{R. Andrae}
\affiliation{Max-Planck-Institut für Astronomie, Königstuhl 17, D-69117 Heidelberg, Germany}
\email[notshow]{eloisa.poggio@inaf.it}

\author[0000-0003-0429-7748]{M. G. Lattanzi}
\affiliation{INAF - Osservatorio Astrofisico di Torino, via Osservatorio 20, 10025 Pino Torinese (TO), Italy}
\email[notshow]{eloisa.poggio@inaf.it}



\begin{abstract}


We explore the three-dimensional structure of a sample of $\sim$ 16000 young giant stars in the Galactic disc out to $\sim$8 kpc in heliocentric distance. This population traces a thin disc with a local vertical scale height of $h_{Z \odot} = 77 \pm 4$ pc, that progressively thickens toward the outer Galaxy with a prominent Galactic flare, rising exponentially with a radial scale length of $h_{fl} =  3.5 \pm 0.3 \, \rm{kpc}$. Our analysis incorporates both the survey selection function and the vertical displacements caused by the Galactic warp and corrugations, which, if neglected, would lead to significant biases in the derived disc scale height. In the Galactic plane, the young giants trace coherent spiral arm segments, extending previous maps based on upper main sequence (UMS) and OB stars by 2-4 kpc depending on the considered direction. The obtained map supports a pitch angle of roughly 20 degrees for the Perseus Arm, and shows that the Local/Orion arm stretches at least 10 kpc in length. Unlike earlier and more local maps based on UMS and OB stars, where the relatively small sampled portion of the Perseus Arm appeared as a short, nearly straight feature, our map reveals it as an extended structure with a gentle curvature, as expected for spiral arms on large scales. In the inner Galaxy, we also identify a new segment likely associated with the Scutum Arm, clearly detached from the Sagittarius–Carina Arm in the fourth Galactic quadrant.

\end{abstract}

\section{Introduction} 

Spiral galaxies display a characteristic thin, flattened disc, whose thickness varies systematically with stellar age and galactocentric distance, due to the system's evolutionary history \citep{Freeman:2002, Nordstrom:2004}. The vertical scale-height of young, metal-rich stars typically does not exceed a few hundred parsecs, reflecting the dynamically cold and rotationally supported nature of their orbits \citep{Bland:2016,Binney:2023}. Older and [$\alpha$/Fe]-enhanced stellar populations on disc-like orbits appear to be more centrally concentrated \citep{Bensby:2011, Bovy:2012} and typically exhibit a larger vertical thickness \citep{VanDerKruit:2011, Rix:2013, Amores:2017, Mazzi:2024}, likely representing the remnants of a dynamically heated early disc \citep{Quinn:1993, Villalobos:2008, Helmi:2018, Belokurov:2018}. 

Numerical simulations indicate that a variety of dynamical mechanisms can increase the vertical scale height of coeval stellar populations with galactocentric radius, thereby producing a flare in the outer disc \citep{Minchev:2015}. Among these mechanisms, satellite–disc interactions have been found to be the dominant contributor \citep{Kazantzidis:2008, Villalobos:2008}, as they can heat and perturb the disc by transferring ordered kinetic energy into random stellar motions \citep{Quinn:1993}. However, even in the absence of environmental influences, disc flaring can arise purely from secular evolution, driven by radial migration induced by spiral arms and/or a central bar \citep{Minchev:2012}. Interestingly, while radial migration flares discs in isolation, it can act to reduce flaring in the presence of satellite interactions \citep{Minchev:2014}. These factors, acting individually or in concert, encode the formation and interaction history of a galaxy, making disc thickness a key observable for understanding galactic evolution \citep{Toth:1992}.

The mechanisms for flaring mentioned above principally effect evolved stellar populations, and help explain the increase of their scale heights over time. Young stellar populations on the other hand principally inherent their scale heights from the gas from which they were recently born. Nevertheless, we would still expect a flare in such a population as the vertical distribution of the gas disc itself, determined by hydrostatic equilibrium, is flared as as a consequence of the declining surface density of the disc with respect to galactocentric radius. 

In our own Galaxy, a pronounced flare has been observed in the outer disc (farther than approximately 8-10 kpc from the Galactic center) using a great variety of tracers, including HI gas \citep{Nakanishi:2003, Levine_flare:2006}, molecular clouds \citep{May:1997}, OB stars \citep{Li:2019}, Red Giant Branch and/or Red Clump stars \citep{LopezCorredoira:2002, Momany:2006, Uppal:2024, Khanna:2025}, supergiants or Gaia general populations \citep{Chrobakova:2022}, and pulsars \citep{Yusifov:2004}. Observational data suggest that the flaring of the Galactic disc is present across stellar populations of different ages, though its amplitude may vary depending on the population considered. On the other hand, when the total stellar population is considered, geometrically defined thin and thick discs \citep[modelled as the sum of two exponentials, ][]{Gilmore:1983} can exhibit a lack of flaring in the total disc, as a natural consequence of the superposition of coeval populations in an inside-out formation scenario \citep{Minchev:2015}. Therefore, mapping the Galactic flare with coeval stellar populations individually can yield valuable insights into its past formation.


In addition to the vertical structure, the in-plane distribution of stars can uncover key features of the disc, including its large-scale spiral pattern \citep[e.g.][]{Vallee:2017a, Vallee:2017b,Shen:2020}. A wide range of studies has mapped the spiral structure using a diverse set of tracers, including masers \citep{Reid:2019}, neutral hydrogen emission \citep{Levine:2006}, radio and optical data of HII regions \citep{Georgelin:1976, Bland-Hawthorn:2002, Russeil:2003}, HI and CO distribution \citep{Soeding:2025}, direct mapping of OB or OBA stars \citep{Pantaleoni:2021, Poggio:2021, Zari:2021, GaiaCollDrimmel:2023, Quintana:2025a, Quintana:2025} and Cepheids \citep{Bobylev:2022, Drimmel:2025}. Significant progress has also been achieved through analyses of stellar kinematics based on Gaia data, possibly linking the observed locii of the spiral arms to dynamical processes in the Galactic disc \citep[e.g.][]{GaiaCollDrimmel:2023, MartinezMedina:2022, Khanna:2023, Denyshchenko:2024, Asano:2024, Funakoshi:2024, Widmark:2024, Widmark:2025,Zari:2025}.


Nevertheless, a global consensus on the geometry of the Galactic spiral arms remains elusive. Notable discrepancies among existing models and maps persist regarding the arms’ positions, number, orientation, pitch angles, and widths. Although some of these differences may arise from intrinsic Galactic properties \citep[the spiral pattern may or may not appear differently when traced by populations of varying ages, see discussions in ][]{Palicio:2023,Palicio:2025,Khanna:2025,Ardevol:2025,Quinn:2026}, numerous discrepancies are presumably driven by the challenging process of reconstructing the three-dimensional spiral structure from our point of view, embedded in the Galactic disc, inevitably dealing with crowding issues and interstellar extinction. Moreover, an accurate and robust mapping of the spiral structure requires a statistically significant number of tracers with precise distance estimates across large portions of the Galactic disc, ideally based on a highly homogeneous survey selection function. It is evident that assembling a sample that simultaneously fulfills all of these requirements is a highly nontrivial undertaking.

In this contribution, we aim to take a step forward in this direction, by mapping the three-dimensional structure of the Galactic disc using a sample of young giant stars from \emph{Gaia} Data Release 3 \citep{GaiaCollVallenari:2023}. Specifically, here we map the flare and the spiral structure. We will not discuss the disc vertical distortions (warp, corrugations and waves), as they were already studied with a very similar sample in our previous work \citep[][hereafter \citetalias{Poggio:2025}]{Poggio:2025}. We will, however, take them into account while modelling the flare, as neglecting them can potentially cause severe biases, as shown in this work. 

The paper is structured as follows: Section \ref{sec:data} is dedicated to the description of the adopted dataset, while Section \ref{sec:flare} and Section \ref{sec:spiral}, respectively, focus on mapping the flare and the spiral structure of the Milky Way. Finally, we discuss our results in Section \ref{sec:discussion} and present our conclusions in Section \ref{sec:conclusion}.

\section{Data} \label{sec:data}

In this Section, we aim to select an all-sky sample of young giant stars from the \gdr{3} catalog \citep[][]{GaiaCollVallenari:2023}. As will be discussed in the following, the approach adopted here is very similar to the one outlined in Section 2 of \citetalias{Poggio:2025}, with the only main difference being that we will now reach fainter apparent magnitudes. The target stars have typical effective temperatures \T\, $\sim 4500 - 5000$ K and surface gravities \g\, of about $\sim$ 1 dex; they cover a portion of the \T - \g\ diagram which is part of the stellar evolutionary stage known as the blue loop \citep[see the massive sample in][]{DR3-DPACP-104}. Here, in particular, we are interested in its cold phases \cite[see Sample A in][]{Poggio:2022}. The comparison with the PARSEC isochrones \citep{Bressan:2012,Chen:2014,Chen:2015,Pastorelli:2019} suggests that, in the range of typical disc metallicities, this portion of the \T - \g\ diagram is expected to be populated by young stars (with ages of about $\sim$100 Myr or less), as shown by Fig. 1 of \citetalias[][]{Poggio:2025}. The kinematic behaviour of these tracers \citep[see Sample A in Fig. B2 in][]{Poggio:2022} and the presence of spiral structure in their density distribution \citepalias[see Fig. 5 in][ and Section \ref{sec:spiral} of this paper]{Poggio:2025} shows that those stars are indeed expected to be young. The target stars are valuable tracers of the young Galactic disc thanks to their typically bright absolute magnitudes (approximately between -3 and -7 magnitudes in the G band), which make them visible out to relatively large distances. Moreover, compared to other young tracers of the Milky Way (e.g. classical Cepheids), they are relatively numerous, enabling a statistically robust analysis of the disc's three-dimensional shape, as will be discussed in the following Sections. Here we will describe the most relevant steps of the sample selection, and refer the reader to Section 2 of \citetalias{Poggio:2025} for more details.

The XGBoost catalog \citep{Andrae:2023} provides stellar parameters (\T, \g, [M/H]) for $\sim$ 175 million stars, based on \gdr{3} XP spectra \citep{DeAngeli:2023, Montegriffo:2023}, combined with CatWISE photometry \citep{Marocco:2021}. From the XGBoost catalog, we select only sources that satisfy the following cuts in the Kiel diagram:
\begin{equation}
\begin{aligned}
&0.5 <  \text{\g} /  \text{dex}  < 2 \\ 
&\text{\g} > ( C \cdot \text{\T} + I_{L} ) \\ 
&\text{\g} < ( C \cdot \text{\T} + I_{R} ) 
\end{aligned}
\end{equation}
where the coefficient C=0.00192 $\rm{ dex \, K}^{-1}$ approximately follows the typical inclination of the Red Giant Branch \citep[using the slope adopted in][ to select their Massive sample]{DR3-DPACP-104}, while the intercepts $I_{L} = -8.3$ dex and $I_{R} = -7.3$ define respectively the left and right borders of the selected area.\footnote{The selected area on the Kiel diagram is indicated in Fig. 1 of \citetalias{Poggio:2025}.} After applying this preliminary selection, we further refine our sample by keeping only stars with apparent magnitude $G<17$, obtaining 544,552 stars. This cut is motivated by the fact that, for magnitudes fainter than $G=17$, the catalog appears to be heavily incomplete. Moreover, as anticipated, this cut is less conservative than the one applied in \citetalias{Poggio:2025} (which only included stars with $G<15$, down to which the catalog appears to be reasonably complete). The choice of including stars fainter than $G=15$ is to allow us to map as large a portion of the Galactic disc as possible. In Section \ref{sec:flare}, we will compare our dataset to a model that takes into account the selection function (and, therefore, the incompleteness of the sample for $G>15$), to avoid the risk of obtaining biased results when exploring the distribution of fainter stars. We note that, however, all the steps of the analysis performed in Section \ref{sec:flare} and \ref{sec:spiral} of this paper will be repeated with the most conservative sample of \citetalias{Poggio:2025}, to verify the robustness of the results presented here. 

Distances to individual sources are calculated as described in \citetalias{Poggio:2025}, and we refer the reader to Section 2.2 and Appendix B of that paper to find the description of the adopted procedure, with the only difference that the sub-selection function was calculated using the apparent magnitude $G<17$ instead of $G<15$, to be self-consistent with the selection adopted in this work.

The preliminary selection described above might have collected not only young stars with disc-like metallicity, but also a fraction of metal-poor contaminants of significantly larger age \citepalias[see more details in Section 2.3 of][]{Poggio:2025}. To get rid of the potential contaminants, we performed some additional cleaning cuts, which we describe in the following. First, we removed stars with metallicities lower than $[M/H]_{lim} = -0.04 \cdot R - 1.25$, where $R$ is the Galactocentric radius \citep[calculated assuming a distance from the Galactic center of $R_{\odot}=8.277$ kpc, following][]{Gravity:2022} and the coefficient $-0.04$ dex/kpc roughly follows the typical values of the metallicity gradient of the Galactic disc \citep[e.g.][]{DR3-DPACP-104}. The intercept has been tuned to perform a broad cut in metallicity, which accounts for the potential scatter induced by metallicity uncertainties and intrinsic metallicity dispersions, but at the same time  minimizes the contamination from metal-poor stars.  

To further clean our sample from potential contaminants, for the fraction of stars that have available radial velocities ($\sim80 \%$ of the sample), we removed stars with low azimuthal velocity $V_{\phi} < 180$ km/s \cite[assuming, for the solar velocity, the values reported in ][]{Drimmel:2018}. We removed probable globular cluster members using the catalog from \cite{Vasilev:2021} (considering the sources with membership probability greater than $50\%$) and \cite{Hunt:2023}. We also removed stars falling in the region of the sky of the Magellanic Clouds, respectively, by removing stars with $275^{\degree}<l<285^{\degree}$ and $-38^{\degree}<b<-28^{\degree}$ (for the Large Magellanic Cloud), and $292^{\degree}<l<310^{\degree}$ and $-47^{\degree}<b<-41^{\degree}$ (for the Small Magellaninc Cloud), with heliocentric distances larger than 3 kpc.

\begin{figure}[t!]
    \centering
    \includegraphics[width=\linewidth]{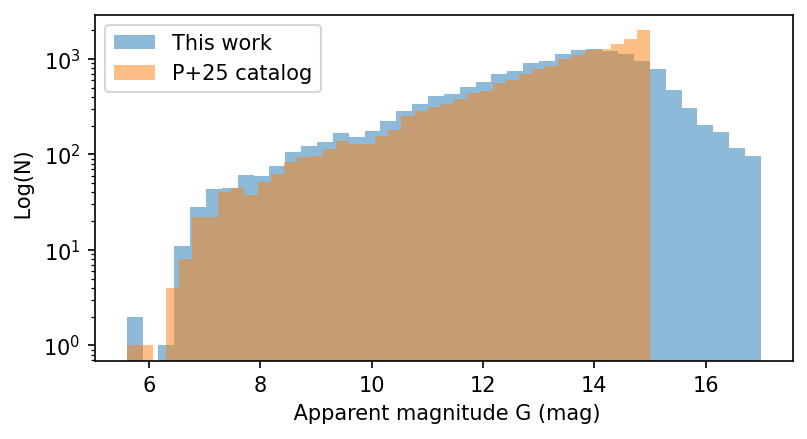}
    \caption{Histogram of the apparent magnitude G for the sample selected in this work (blue) and the young giant catalog from \citetalias{Poggio:2025} (orange).
\label{fig:histogram_G}}
\end{figure}

Unlike \citetalias[][]{Poggio:2025}, we do not apply a galactic latitude cut, to avoid an artificial hole close to the Sun when creating X-Y maps of the stellar distribution. Finally, we also apply a cut on parallax relative error $\varpi / \sigma_{\varpi} < 5$, which helps us to remove the stars with poor astrometric quality at fainter magnitudes, which would lead to unreliable distances. The final catalog contains $16\,348$ stars. We note that the above-mentioned parallax cut was not necessary in the catalog \citetalias[][]{Poggio:2025}, as the overall astrometric quality of the data was already enough to obtain reliable distances, being limited to $G<15$ mag. The sample selected here and the one from \citetalias[][]{Poggio:2025} can be, therefore, seen as complementary. Their apparent magnitude distribution is shown in Figure \ref{fig:histogram_G}.

\section{Flare} \label{sec:flare}

\subsection{Overview}

\begin{figure*}[ht!]
\centering
\includegraphics[width=0.99\textwidth]{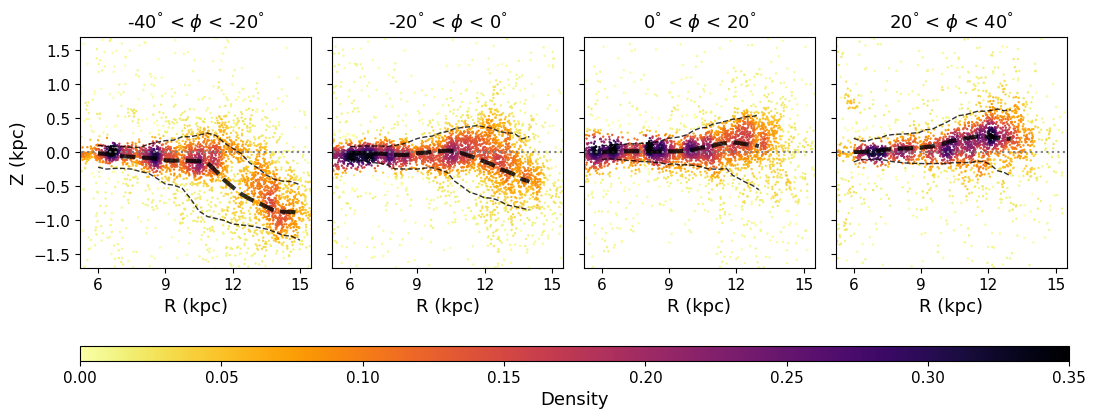}
\caption{Edge-on view of the Galactic disc at different azimuthal slices. The Sun's position is at Galactic azimuth $\phi=0\degree$ and $\phi>0\degree$ in the direction of Galactic rotation. The thick dashed line shows the median vertical distribution at a given Galactocentric radius R within $\pm 1 \, \rm{kpc}$, while the thin dotted lines show the 84$^{th}$ and 16$^{th}$ percentiles of the vertical distribution. The inhomogeneous clumps in the density are cross-sections of the spiral arms, as discussed in the text.
\label{fig:overview_flare}}
\end{figure*}

In this Section, we analyze the flare of the Milky Way disc using the dataset selected in Section \ref{sec:data}. Figure \ref{fig:overview_flare} shows the vertical distribution of our dataset along the Z-axis as a function of Galactocentric radius $R$, for different bins in Galactic azimuth $\phi$. In our coordinate system, $\phi=0^{\degree}$ coincides with the line that connects the Galactic center, the Sun's position and the Galactic anticenter. The azimuthal angle $\phi$ is taken as positive in the direction of Galactic rotation. The four different panels of Figure \ref{fig:overview_flare} show that, in the outer parts of the disc, the vertical position of the midplane (i.e. thick black dashed line) deviates from $Z=0$, typically downwards for approximately $\phi<0^{\degree}$, and upwards for approximately $\phi>0^{\degree}$. The observed vertical distortion is expected, as the disc of the Milky Way is known to be warped and corrugated, and several observations show that the Sun ($\phi=0^{\degree}$) lies relatively close to the warp's line-of-nodes. The vertical distortion of the Galactic disc has been extensively studied in \citetalias{Poggio:2025} with a sample very similar to the one adopted in this work, and therefore will not be specifically studied in this contribution. We note, however, that another aspect stands out from Figure \ref{fig:overview_flare}: the typical thickness of the disc (visually shown by the separation between the two thin dashed lines at a given Galactocentric radius $R$) tends to be larger in the outer disc. This feature is known as the Galactic flare, and, in this Section, we aim to characterize its properties with our selected sample. Finally, we note that the density distribution of sources does not decrease monotonically with $R$. Instead, it exhibits several distinct concentrations that correspond to the vertical cross-sections of the Galactic spiral arms. This suggests that we can use our sample also to explore the spiral structure of the Galactic disc, which will be done in Section \ref{sec:spiral}.

\subsection{Methodology}

Here we describe the methodology adopted to model the Galactic flare using our sample. The first step of our approach is to construct 2-dimensional cells in the Galactic plane, using a heliocentric Cartesian coordinate system with a vertical axis perpendicular to the Galactic disc, and the XY-plane aligned with the plane of the Galaxy (at galactic latitude $b=0\degree$). For each cell, the vertical scale-height $h_z$ is inferred, following the procedure that will be explained in the following paragraphs. The chosen bin width along the X and Y axis must be large enough to ensure a sufficient number of stars in the cell (so that the scale-height $h_z$ can be robustly inferred), but, at the same time, small enough to allow us to map large-scale variations of the the vertical scale-height $h_z$ in the Galactic disc, as well as accounting for the systematic vertical variation of the midplane from the warp and corrugations (see below). After performing several tests, we find that a bin width of 1.5 kpc in X and Y represents a good compromise between these two aspects. 

As can be immediately seen from Figure \ref{fig:overview_flare}, inferring the disc scale height at a given position without rescaling the entire vertical distribution for the median $Z_{MED}$ (thick dashed lines) can potentially lead to severe biases, especially in the outer Galaxy. Indeed, while in the inner disc $Z_{MED}$ is typically very close to zero, it can significantly deviate from the $Z=0$ plane in the outer parts, due to the presence of the Galactic warp and vertical corrugations. Therefore, for each XY-cell, we rescale the vertical coordinates to the median vertical height of the stars within that cell ($Z_{MED}$) in order to account for this effect: 
\begin{equation}
\label{eq:vertical_rescaling}
Z^{\prime} = Z-Z_{MED} \quad,
\end{equation}
where $Z= d \sin{b} + Z_{\odot}$, with the heliocentric distance $d$ obtained in Section \ref{sec:data}, and $Z_{\odot}$ is the Sun's height above the plane.
Since we are rescaling the vertical distribution to $Z_{MED}$, we assume $Z_{\odot}=0 \, pc$ for convenience.
To determine the vertical scale-height $h_z$ within each cell, we adopt a method we refer to as the \emph{$Z^{\prime}$-binning} approach, which is described below in Subsection \ref{subsec:Zprimebinning}. We first verify the reliability of our approach using mock catalogs (Section \ref{subsec:mock_catalog}), and finally apply them to real data (Section \ref{sec:results}). Once $h_z$ is obtained for each cell of our dataset, we explore its variation in the Galactic disc, and parameterize its increase as a function of Galactocentric radius $R$ (Section \ref{sec:results}).

As an additional test, we also adopt an alternative to the \emph{$Z^{\prime}$-binning} approach that is based on the likelihood of individual stars, hereafter referred to as \emph{Individual likelihood} approach. A detailed description of this method, including its application to both mock and real data, can be found in Appendix \ref{subsec:individual_likelihood}.

\subsubsection{$Z^{\prime}$-binning approach}\label{subsec:Zprimebinning}

In this Section, we will describe the \emph{$Z^{\prime}$-binning} method. This approach requires binning the stars in each $XY$ cell into subcells along the $Z^{\prime}$-axis, following the rescaling defined in Equation \ref{eq:vertical_rescaling}. The binning size in $Z^{\prime}$ is chosen following the Freedman-Diaconis rule \citep{Freedman:1981}. However, to ensure that the obtained results are not strongly influenced by the adopted binning size, alternative strategies have been tested, like adopting a fixed number of bins (e.g. 10, 15 or 20 bins) within 3-$\sigma$ of the distribution in $Z^{\prime}$, or, alternatively, following Scott's rule \citep{Scott:1979}. For each cell, we infer $h_z$ with the \emph{emcee} package \citep{Foreman-Mackey:2013} using the Poissonian\footnote{A Gaussian likelihood has also been tested.} likelihood function
%
\begin{equation}
\label{eq:binning_fit}
\mathcal{L} = \prod_i \frac{N_{pred,i}^{N_i}}{N_i!}e^{-N_{pred,i}} \quad,
\end{equation}
for which, for convenience, we adopt the logarithmic form:
\begin{equation}
\log\mathcal{L} = \textrm{const} + 
\sum_i \left(N_i\cdot\log(N_{pred,i})-N_{pred,i}\right) \quad,
\end{equation}
where $N_i$ is the observed number of stars for a given $Z^{\prime}$ bin, $\sigma_i = \sqrt{N_i}$ is the corresponding uncertainty (assuming Poissonian noise), and $N_{pred,i}$ is the number of stars predicted by our model at that specific bin. As shown below, $N_{pred,i}$ represents the number of observable objects, and will not only depend on the intrinsic scale height of the Galaxy, but also on the extinction and selection function of our dataset. Specifically, we can write
\begin{equation}
\label{eq:observable_stars}
N_{pred,i} = {\mathcal{F}}_{i} \cdot N_{disc,i}  \quad,
\end{equation}
where $N_{disc,i}$ is the number of stars predicted by the disc model at that specific location, and ${\mathcal{F}}_{i}$ is the fraction of stars that we can observe, taking extinction and the selection function into account. For each XY-cell, the number density of (thin) disc stars at a given $Z^{\prime}$ can be written as 
\begin{equation}
\label{eq:exp_profile_cell_nfloor}
N_{disc,i}(Z^{\prime}) = N_{floor} +  N_0 \exp{ \left( - \frac{|Z^{\prime}|}{h_{Z}} \right)} \quad,
\end{equation}
where $h_{Z}$ is the disc scale height at that specific XY-position, $N_0$ is the normalization factor, and the $N_{floor}$ parameter takes into account the possible presence of contaminants. The three parameters are constant within each cell, but are permitted to vary across different cells in the Galactic disc. While the fraction of potential contaminants is expected to be very low (as will be confirmed in Section \ref{sec:results}), the choice of using the $N_{floor}$ parameter is to insure that our inferred scale height is not biased. Testing with a model without the $N_{floor}$ parameter showed it to be particularly beneficial in two specific cells towards the Galactic center. 


To compute the fraction of observable stars ${\mathcal{F}}_{i}$, we first calculate the luminosity function $\Phi(M_G)$, where $M_G$ is the absolute magnitude in the G band, using the PARSEC isochrones \citep{Bressan:2012, Chen:2014, Chen:2015, Pastorelli:2019}, and applying the same \text{\T}-\text{\g} cuts that were applied to select the data in Section \ref{sec:data}. More details on the calculation of the luminosity function can be found in \citetalias{Poggio:2025}. We then convert the coordinates of each subcell $(X,Y,Z)$ into heliocentric coordinates $(l,b,d)$, so that we can express the fraction of observable stars as
\begin{equation}
{\mathcal{F}}_{i}(l,b,d) = \int_{G_{min}}^{G_{max}} S(l,b,G)\, \Phi(M_G (G, d)) \, dG
\end{equation} 
where $G$ is the apparent magnitude, $S(l,b,G)$ is the selection function of our selected sample and $M_G = G - 5 \log(d) + 5 + A_G(l,b,d)$, where $d$ is in parsecs, and $A_G$ is the extinction in the G band, obtained from the \emph{mwdust} package \citep{Bovy:2016}. 
The adopted limits of integration are $G_{max}=17 \, \rm{mag}$ and $G_{min}=5$, outside of which our catalog contains no stars.
Following the approach described in \citet[][]{Castro-Ginard:2023}, we estimate $S(l,b,G)$ by computing the ratio of stars included in the catalogue of \citet{Andrae:2023} to the entire \gaia\ catalogue in bins of $(l,b,G)$, between $4<G<17$. On the \gaia\ archive\footnote{\url{https://gea.esac.esa.int/archive/}} we compute this ratio in bins of $G$ (1 mag wide) and sky position ($l,b$) pixelized at \textit{HEALPix} level 5 \citep{HEALPix2005}. Using the obtained number counts, we calculate $S(l,b,G)=\frac{k+1}{n +2}$ as the mean probability for a source to end up in our subsample, where $n$ is the number of stars in the entire \gaia\ catalogue that fall within our chosen magnitude range with \gaia\ \& WISE photometry, and $k$ is then the set of stars in the sub-sample for which XGBoost parameters are available (those with published \gaia\ DR3 XP spectra). More details can be found in Appendix B of \citetalias{Poggio:2025}. 

For each XY-cell, the $16^{th}$, $50^{th}$ and $84^{th}$ percentiles of the $h_z$ posterior distribution are taken, respectively, as lower uncertainty, best-fit value and upper uncertainty of the estimated scale-height, based on the likelihood outlined in Equation \ref{eq:binning_fit}, and a flat uninformative prior on $h_z$, only imposing the condition that the disc scale-height must be positive.


\subsubsection{Test with mock catalogs}\label{subsec:mock_catalog}

To validate the approach described in Section \ref{subsec:Zprimebinning}, we construct realistic mock catalogs with a known scale height $h_{Z,TRUE}$, and verify that we are able to recover the true value by applying our method. To construct a mock catalog, we first generate the \emph{true} vertical distribution of our mock stars by drawing random samples from an exponential distribution (see Equation \ref{eq:exp_profile_cell_nfloor}, with and without $N_{floor}$ parameter) with scale height $h_{Z,TRUE}=0.1$ kpc. For simplicity, the median Z coordinate is assumed to be 0, so that, in this case, $Z=Z'$. To mimic the stellar distribution of stars in a given cell, we assign (for simplicity) uniform random values of X and Y coordinates in the Galactic disc, within a range of 1.5 kpc in width, to mimic the size of cells in real data. Once the three-dimensional distribution is constructed, we calculate the \emph{true} distance to each mock star, and its corresponding \emph{true} parallax. We then assign absolute magnitudes $M_G$, following the modeled luminosity function of the young giants described in \citetalias{Poggio:2025}; moreover, we apply the {\em mwdust} extinction map from \citet{Bovy:2016} to obtain the apparent magnitudes $G$. We simulate formal parallax uncertainties by assigning the values given by PyGaia\footnote{\url{https://pygaia.readthedocs.io/en/latest/index.html}} as a function of apparent magnitude $G$, with parallax formal uncertainties increasing for fainter stars. We then generate mock observed parallaxes by applying to each true parallax a random error, drawn from a Gaussian distribution with standard deviation equal to the formal parallax uncertainty. We obtain the mock observed distances by inverting the mock observed parallaxes, and then obtain the Z-coordinate. Only the observed quantities are then used to infer the intrinsic disc scale height  $h_{Z,TRUE}$ using our methodology, as done with the real data. 
We note that the methodology adopted in this work to obtain distances to individual stars with real data is expected to be significantly more reliable than the simple process of inverting the observed parallax \citep[as shown for example by ][]{BailerJones:2015,Luri:2018}, which is here adopted to construct the mock catalog for the purpose of simplification. Therefore, the results obtained with the mock catalog are presumably more prone to distance errors than the real data should be, leading to a conservative approach. As a final step, we remove from the mock catalog the stars that are not expected to be observable, based on our survey selection function (modelled in the previous Sections), depending on their mock sky coordinates (l,b) and apparent magnitude G.

To test the reliability of the $Z'$-binning approach, we generate 10 mock catalogs with $N_{mock}$ stars to infer the vertical thickness. For a given cell, if we construct mock catalogs with $N_{mock} \sim 200$, we find that the true value $h_{Z,TRUE}$ is recovered within one $\sigma$ in 60$\%$ of the cases, and within two $\sigma$ in 100$\%$ of the cases for a typical heliocentric distance of 4 kpc. If we increase $N_{mock}$ to 300-400 stars, we tend to obtain more precise results, with the true value $h_{Z,TRUE}$ recovered within one $\sigma$ for 70$\%$ of the cases, and within $1.2\sigma$ in 100$\%$ of the cases. If we increase the typical distance of the cell to about 6 kpc, we find that, for $N_{mock} \sim 200$, $h_{Z,TRUE}$ is successfully recovered within one $\sigma$ in 70$\%$ of the cases, and within $2\sigma$ in 100$\%$ of the cases. We note that, for the most distant cells, the relative apparent size of the cell on the sky is smaller, due to geometrical projection effects. As a consequence, the modelization of the extinction and selection function is presumably more effective for the distant cells than the nearby ones. While distance errors tend to increase with distance (as expected), this effect might be less crucial than the extinction/selection function for the inference of $h_{Z,TRUE}$, making the inference process successful even for the most distant cells. To summarize our results, the test performed with our mock catalogs reveals that the $Z'$-binning approach successfully recovers the true scale height $h_{Z,TRUE}$ within one $\sigma$ in most cases, and always within $2\sigma$.

\begin{figure*}[ht!]
\centering
\includegraphics[width=0.33\textwidth]{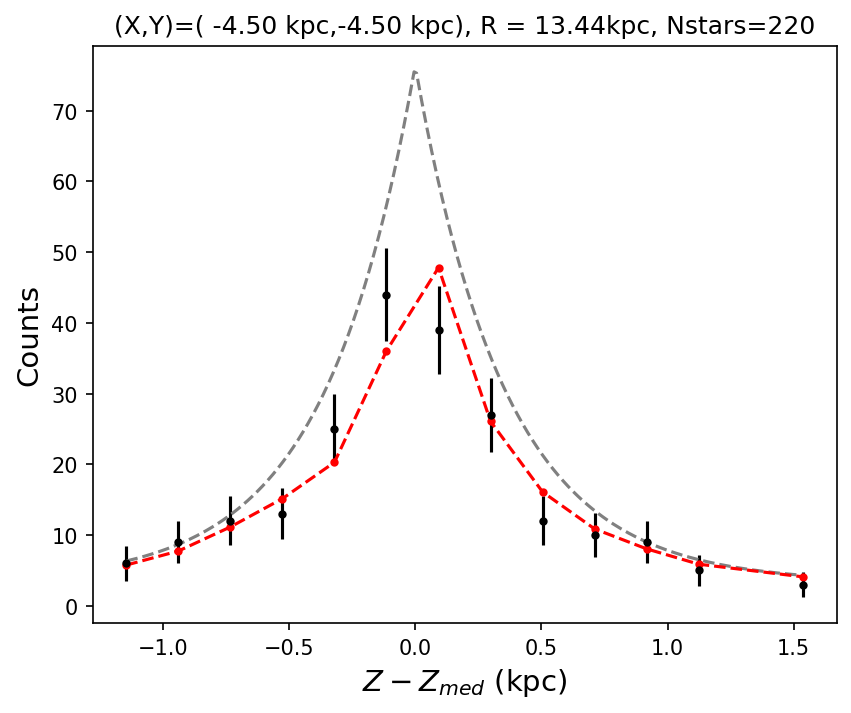}
\includegraphics[width=0.33\textwidth]{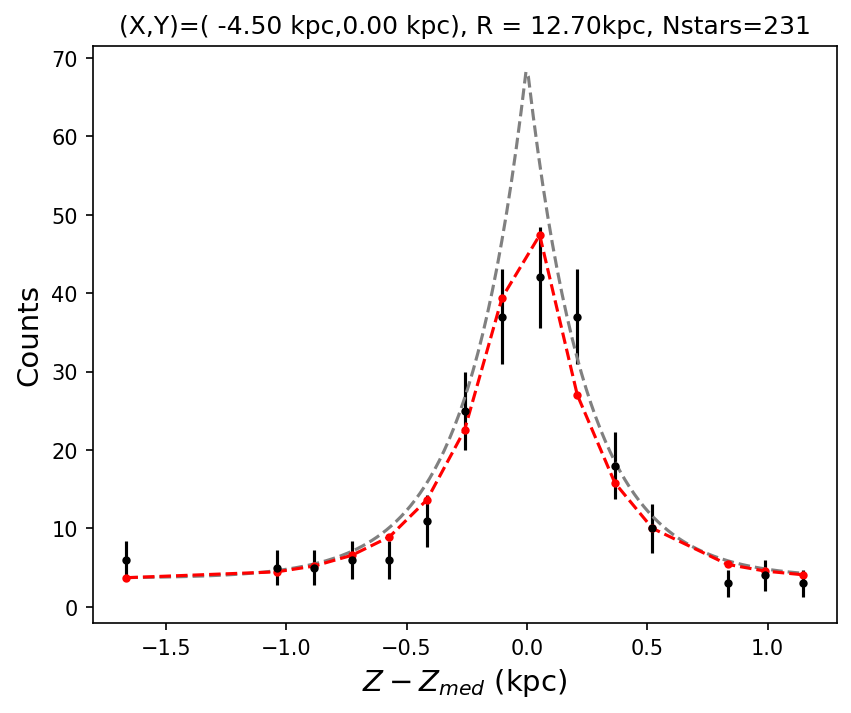}
\includegraphics[width=0.33\textwidth]{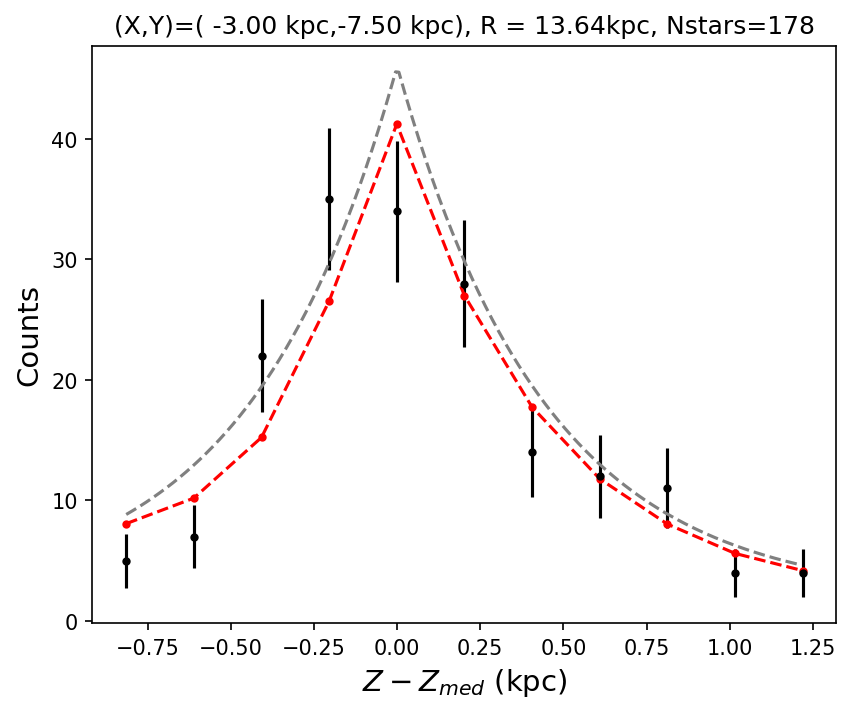}
\includegraphics[width=0.33\textwidth]{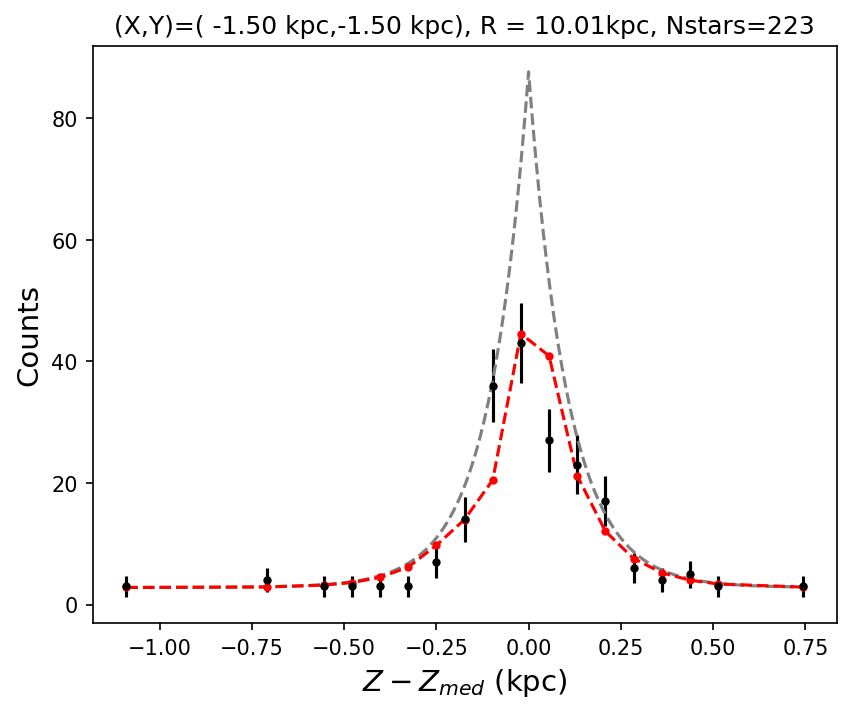}
\includegraphics[width=0.33\textwidth]{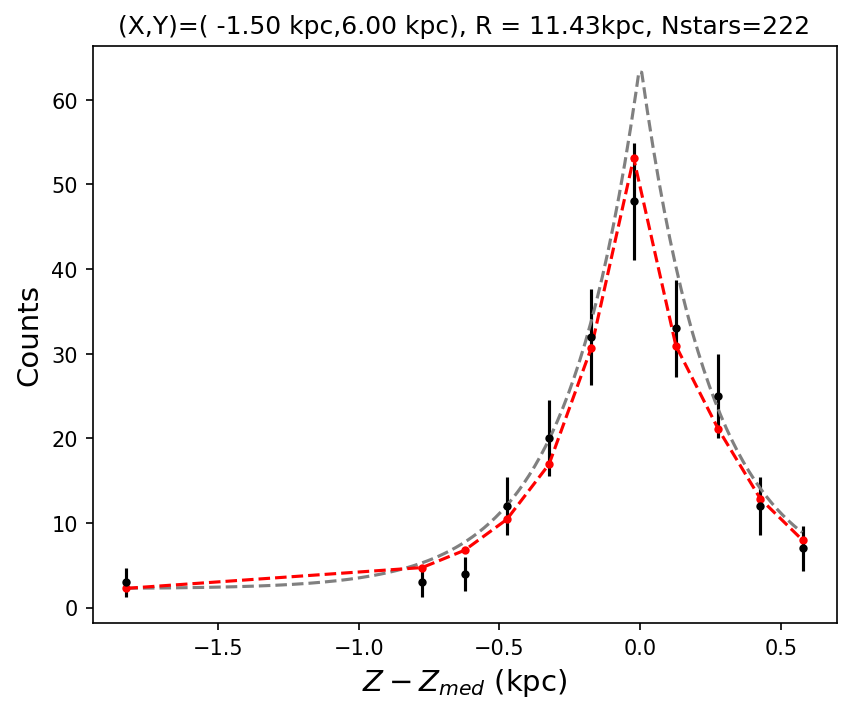}
\includegraphics[width=0.33\textwidth]{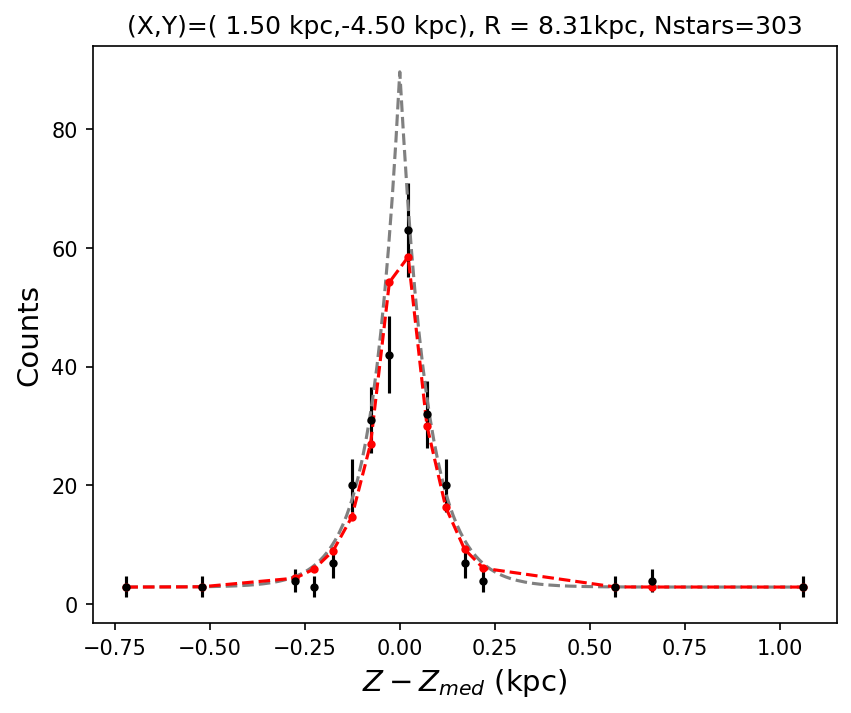}
\includegraphics[width=0.33\textwidth]{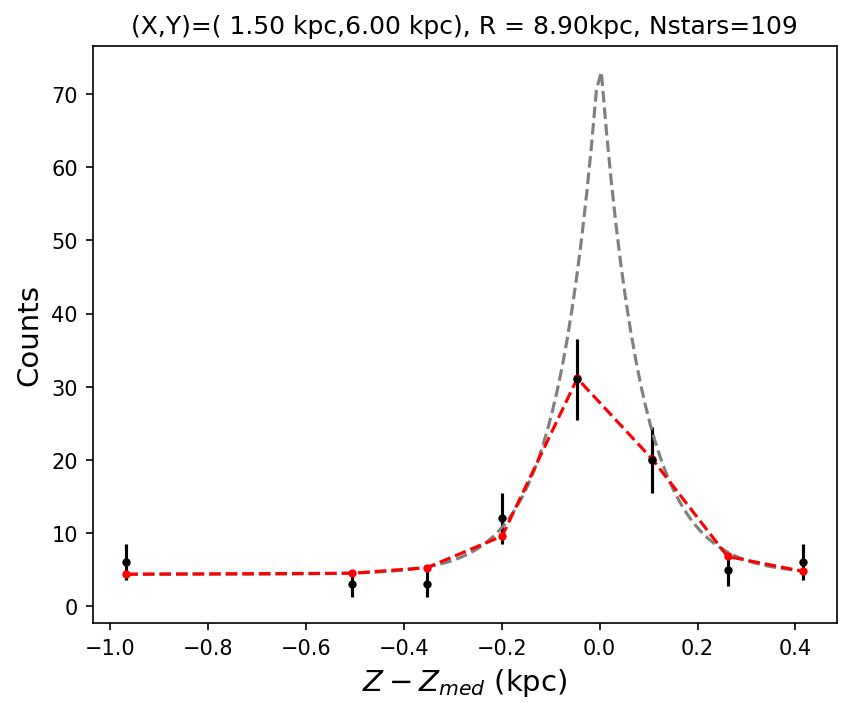}
\includegraphics[width=0.33\textwidth]{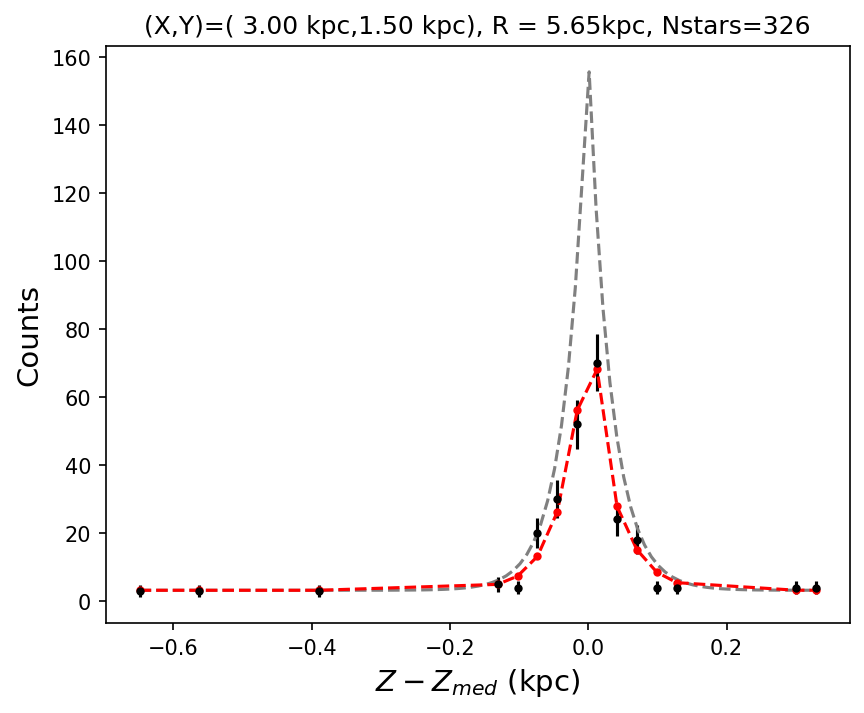}
\includegraphics[width=0.33\textwidth]{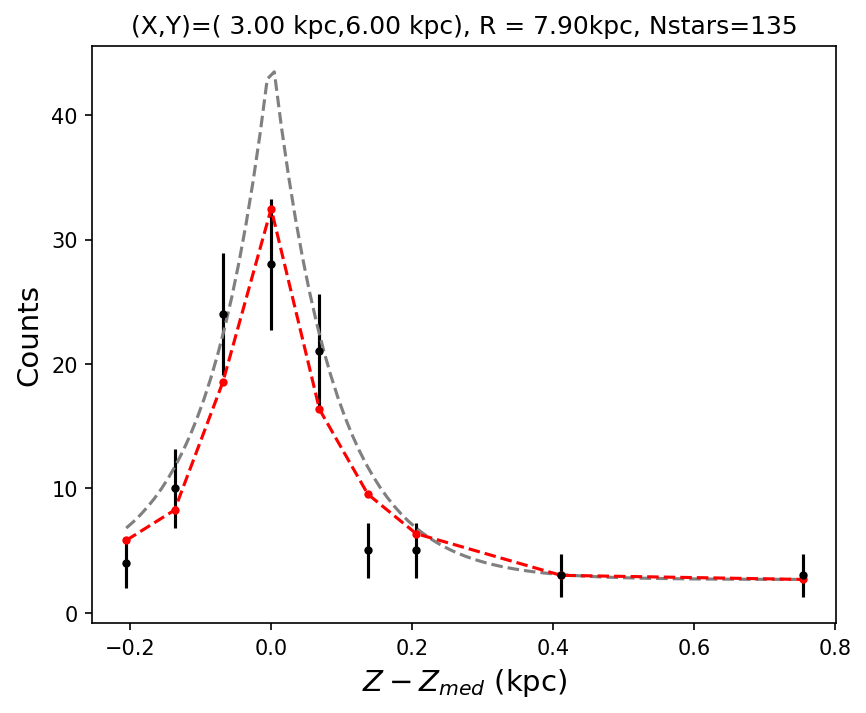}
\caption{Examples of fitted vertical distributions in individual cells. Each panel corresponds to a different cell in the XY-plane. For a given cell, the black points show the observed star counts along the vertical coordinate $Z^{\prime} = Z - Z_{med}$, where $Z_{med}$ is the median Z of the stars in that cell. The red dashed line shows the best-fit model obtained in this work. For illustrative purposes only, the grey dashed line shows same best-fit model, but without accounting for the survey selection function and extinction.
\label{fig:examples_fits_cells_hz}}
\end{figure*}

\subsection{Results}\label{sec:results}

\begin{figure*}[ht!]
\centering
\includegraphics[width=\textwidth]{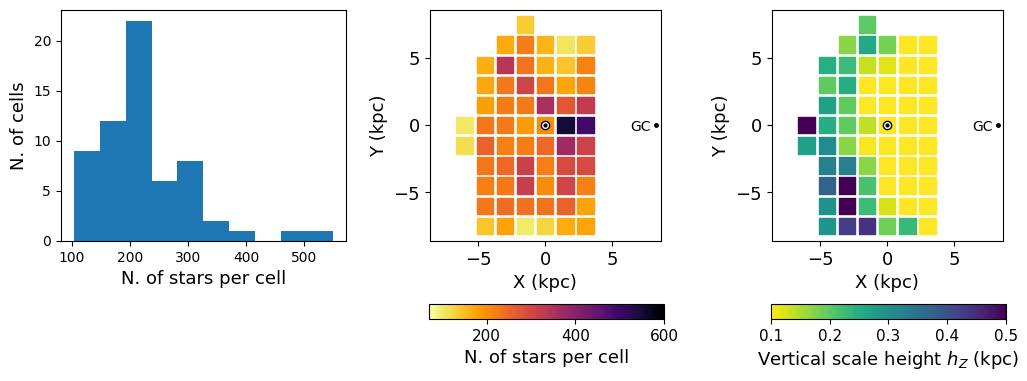}
\caption{\emph{Left panel}: histogram of the number of stars per cell. \emph{Middle panel}: distribution of the cells in the XY-plane, color coded by the number of stars per cell. The Sun's position is in (X,Y)=(0, 0), and the Galactic center is to the right, in (X,Y)=($R_{\odot}$,0), with $R_{\odot}=8.277$ kpc \citep{Gravity:2022}. \emph{Right panel}: same as the middle panel, but now color-coded by the inferred $h_Z$ in each cell.
\label{fig:xy_map_hz_results_fits_cells}}
\end{figure*}

Figure \ref{fig:examples_fits_cells_hz} shows some examples of the observed distribution of stars along the coordinate $Z'=Z-Z_{MED}$, where $Z_{MED}$ is the median $Z$ of the stars in the cell, together with their best-fit model (red dashed lines). The corresponding model obtained without considering extinction or the selection function is also plotted (grey dashed lines) for illustrative purposes only. As expected, we find that the fit quality depends on a number of factors, including the total number of stars in a given cell and the quality of their estimated distances. We apply a minimum threshold of 100 stars per cell, resulting in a total of 62 cells. The selected cells are generally confined to within approximately 7 kpc of the Sun, depending on the considered direction. Cells with fewer stars tend to have larger distances and large relative errors on $h_z$, which means that they don't have enough data to reliably constrain the vertical thickness, leading to noisy measurements. However, as will be discussed below, the decision to impose or not impose a minimum threshold on the number of stars has no effect on the final flare parameters, given that identical values are obtained in both cases. As we can see from the left panel of Figure \ref{fig:xy_map_hz_results_fits_cells}, the typical number of stars per cell is between 100 and 350, but there are a few exceptions with larger numbers. The individual XY-position of each cell in the Galactic disc is shown in the middle and right panel of Figure \ref{fig:xy_map_hz_results_fits_cells}, color-coded by the number of stars in each cell and the inferred $h_{z,obs}$, respectively. The cells were constructed to provide contiguous, non-overlapping coverage of the disc. This design choice ensures that the vertical scale-heights obtained in each cell are statistically independent, which will be crucial in the following part of our analysis.

\begin{figure}
    \centering
    \includegraphics[width=\linewidth]{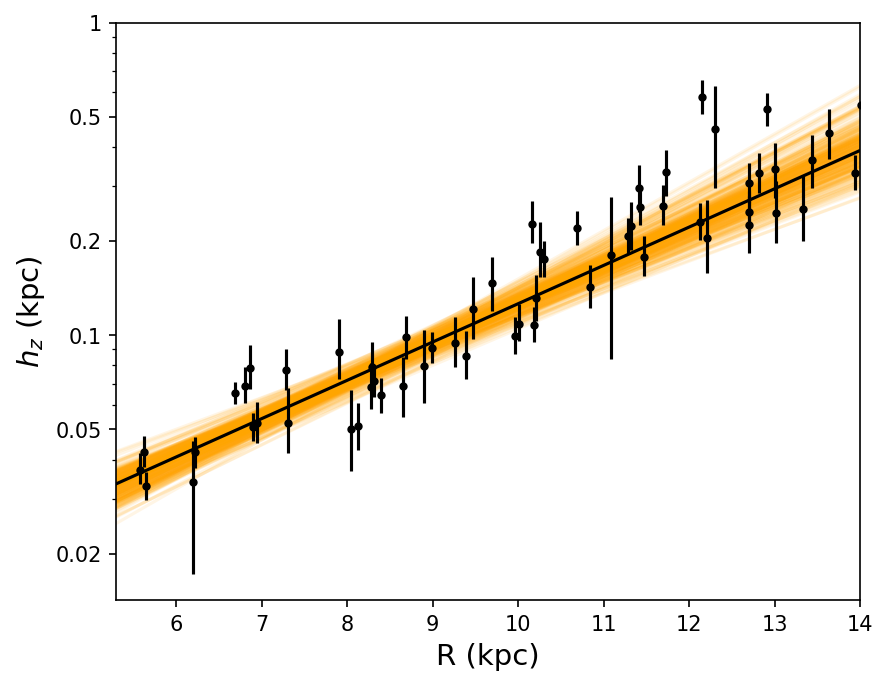}
    \caption{Obtained disc scale heights at different Galactocentric radii R. Each point corresponds to a cell in the middle and right panels of Figure \ref{fig:xy_map_hz_results_fits_cells}. The black solid line shows the best-fit exponential model of the Galactic flare, and the transparent orange lines are obtained by drawing random samples from the uncertainties associated with $h_{z0}$ and $h_{fl}$, as explained in the text. Given that the y-axis is on a logarithmic scale, the exponential form adopted in Equation \ref{eq:flare} appears as a linear function.}
    \label{fig:hz_vs_R}
\end{figure}

We now explore how the values of vertical scale height $h_{z,obs}$ obtained in each non-overlapping cell vary as a function of Galactocentric radius $R$. As we can see from Figure \ref{fig:hz_vs_R}, the obtained values of $h_{z,obs}$ tend to increase systematically with $R$, which is the signature expected from a flared disc. Following several works in the literature \citep[e.g.][]{LopezCorredoira:2002, Li:2019, Khanna:2025}, we adopt an exponential model to describe the Galactic flare: 
\begin{equation}
\label{eq:flare}
h_z(R) = h_{Z0} \exp{(R-R_{\odot})/h_{fl}}\quad,
\end{equation}
where $h_{Z0}$ is the disc scale height at the Sun's position and $h_{fl}$ is the typical flare scale length. Using the individual values of vertical thickness and the corresponding uncertanities previously obtained in each cell, we infer the parameters $h_{Z0}$ and $R_{fl}$ using the \emph{emcee} package \citep{Foreman-Mackey:2013}. We write the likelihood function as follows:
\begin{equation}
\label{eq:likelihood_fit_cells}
\mathcal{L} = \Pi_{j} \frac{1}{\sqrt{2 \pi} \, \sigma_{h_{z},j}} \exp{- \frac{(h_{z,obs,j} - h_{z}(R_{cell,j}))^2}{2 \, \sigma_{h_{z},j}^2}  } \quad,
\end{equation}
where, for the $j^{th}$ cell, $h_{z,obs,j}$ is the obtained vertical thickness, $\sigma_{h_{z},j}$ is its uncertainty, and $h_{z}(R_{cell,j})$ is the vertical scale height of the disc predicted by the model of Equation \ref{eq:flare} at the Galactocentric radius of the cell. The product $\Pi_{j}$ in Equation \ref{eq:likelihood_fit_cells} is over all the considered cells in the XY plane. As shown in Figure \ref{fig:hz_vs_R}, the observed increase of $h_{z,obs}$ with $R$ is well-characterized by an exponential law. This behavior arises spontaneously from the data, as the individual cells are statistically independent. Therefore, our analysis confirms the exponential behavior reported or assumed in previous literature, specifically for the young population analyzed in this work.

Following this procedure, we obtain $h_{Z0} = 77 \pm 2$ pc and $h_{fl} = 3.5 \pm 0.1$ kpc, where the quoted statistical uncertainties are the outputs of the \emph{emcee} fitting routine. The entire analysis was repeated using 100 bootstrap resamples of our dataset, obtaining a standard deviation of 4 pc on the local scale height $h_{Z0}$, and of $0.3$ kpc on the flare radial scale lenght $h_{fl}$, so that our final results for the flare parameters based on our sample are:

\begin{equation}
\begin{aligned}
\label{eq:fit_flare_results}
h_{z0} &= 77 \pm 4 \, \rm{pc} \\
h_{fl} &= 3.5 \pm 0.3 \, \rm{kpc}. \ 
\end{aligned}
\end{equation}

As expected, we find that the relative fraction of potential contaminants ($N_{floor} / N_{tot}$, where $N_{tot}$ is the total number of stars in each cell) is very small, remaining below 5$\%$ in more than 95$\%$ of the cells. The $N_{floor}$ parameter reaches slightly larger values only in a few distant cells, for which the relative uncertainty on $h_z$ is typically large (due to the lower number of objects), and therefore have a very low constraining power in the fit of the flare. Indeed, the final obtained values of $h_{Z0}$ and $h_{fl}$ remain identical, regardless of whether we include the cells with low number counts ($<$100 stars) and/or large relative uncertainties on the inferred $h_z$ ($>$ 40 $\%$).

Finally, to improve the robustness of our results, we repeat the analysis using the young giant sample from \citetalias{Poggio:2025}, obtaining $h_{Z0} = 70 \pm 2$ pc and $h_{fl} = 3.3 \pm 0.1$ kpc. Those results are consistent with those presented in Equation \ref{eq:fit_flare_results} within 2 and 1 $\sigma$, for $h_{Z0}$ and $h_{fl}$.

As a complementary test, we re-perform the inference of $h_z$ in each cell without rescaling the Z-coordinates for the median Z value, to test the potential bias caused by not taking the warp and vertical corrugations of the disc into account. As expected, the obtained $h_z$ in the outer disc are clearly biased toward larger values (reaching values of 1-1.5 kpc) and with much larger relative errors. On the other hand, the impact at the Sun's position is less significant, due to the relatively small amplitude of the disc vertical distortions in the inner Galactic disc.

\begin{figure}
    \centering
    \includegraphics[width=\linewidth]{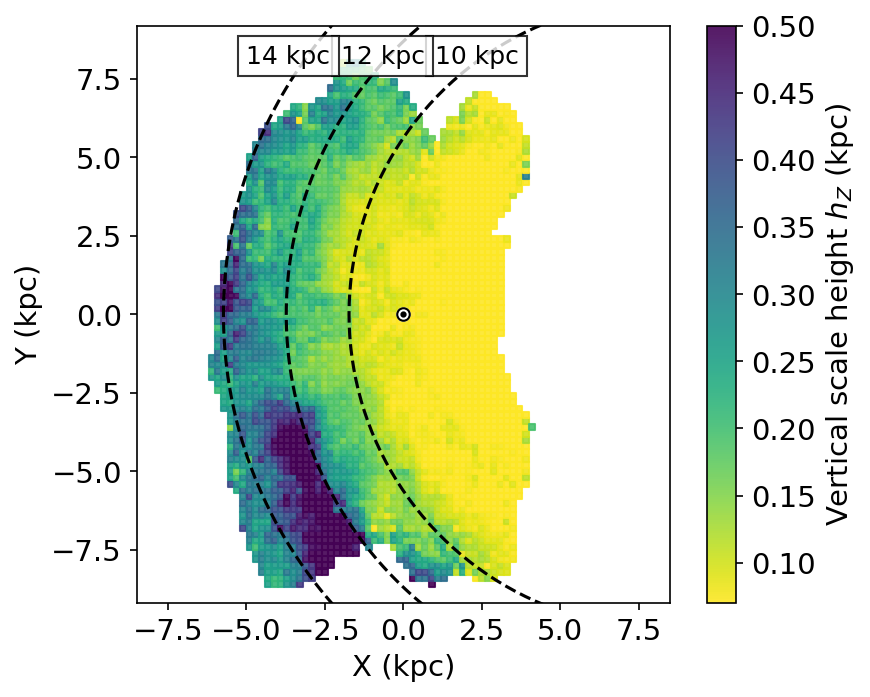}
    \caption{Vertical thickness of the Galactic disc using overlapping cells in XY. The Sun's position is shown by the black cross. Dashed curves represent the constant distance to the Galactic centre at R = 14 kpc, 12 kpc, and 10 kpc, respectively, from left to right.  }
    \label{fig:hz_in_xyplane_overlapping_cells}
\end{figure}

For visualization purposes only, Figure \ref{fig:hz_in_xyplane_overlapping_cells} shows a map of the obtained vertical scale height in the Galactic plane using cells overlapping in XY-space (as opposed to the right panel of Figure \ref{fig:xy_map_hz_results_fits_cells}, which was based on non-overlapping cells). As previously discussed, we find that the vertical scale height tends to increase towards the outer parts of the disc, as expected from a flared disc. We note, however, some azimuthal variation. In particular, between $R \sim 11.5 - 13$ kpc the increase of vertical scale-height as a function of Galactocentric radius tends to be larger in the lower parts of the plot (third Galactic quadrant with heliocentric coordinates $X,Y<0$) compared to the the upper parts (second Galactic quadrant with $Y>0$ and $X<0$). This slight asymmetry can be seen also in the right panel of Figure \ref{fig:xy_map_hz_results_fits_cells} and will be discussed in Section \ref{sec:discussion}.

\section{Spiral structure} \label{sec:spiral}

\begin{figure*}
    \centering
    \includegraphics[width=1\linewidth]{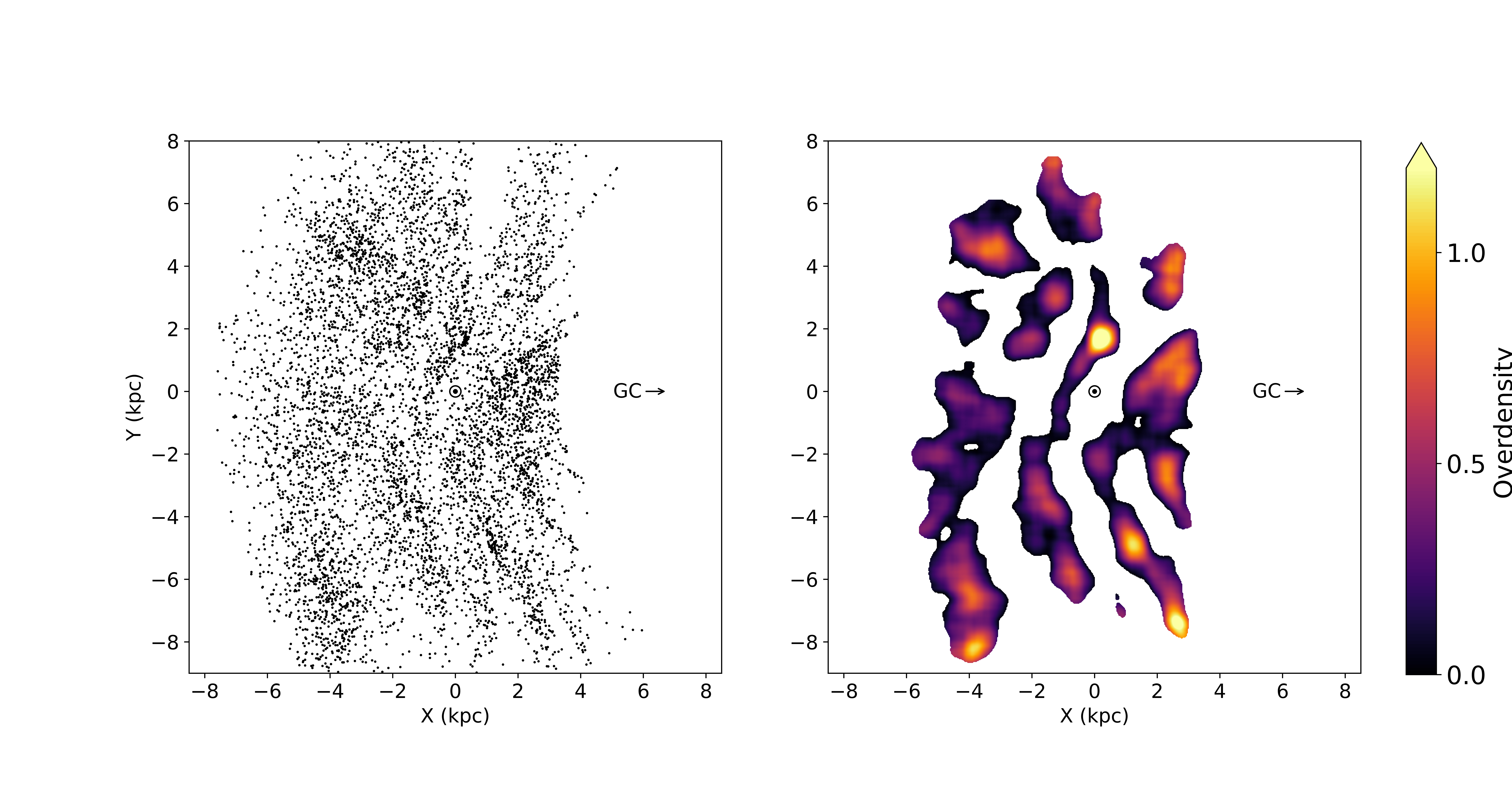}
    \caption{\emph{Left panel}: The distribution of the young giant sample in the XY plane. Each black point corresponds to a star. The Sun's position is shown by the $\odot$ symbol. The Galactic Center (GC) is to the right, at (X,Y)=($R_{\odot}$, 0 kpc), with $R_{\odot}=8.277$ kpc \citep{Gravity:2022}. Galactic rotation is clockwise. \emph{Right panel}: Same as left panel, but now showing the regions of positive ($>$0) overdensities in the Galactic disc based on the young giant sample (see details in the text).}
    \label{fig:spiral_structure}
\end{figure*}

In this Section, we aim to map the spiral structure of the Galactic disc, as traced by our sample of young giant stars. To this end, we perform a vertical cut, keeping only stars at low vertical heights $Z$ above/below the Galactic plane, i.e. those that satisfy the following equation:
\begin{equation}
\label{eq:vertical_cut}
|Z-Z_{MED}| <  0.8 \cdot h_{z}\quad,
\end{equation}
where $h_{z}=h_{z}(R)$ is the flared vertical height of the Galactic disc given by Equation \ref{eq:flare}, using the best-fit parameters obtained in the previous section (Eq. \ref{eq:fit_flare_results}), and the median vertical height $Z_{MED}$ is calculated using all stars within a cell of $\pm 1.5$ kpc in the XY plane centered on each star. The cut applied in Equation \ref{eq:vertical_cut} aims to specifically select the stars located in the Galactic plane, taking into account the vertical distortions (warp, corrugations) and the flare of the young Galactic disc. An assessment of how different vertical cuts influence the present analysis can be found in Appendix \ref{appendix:impact_verticalcuts_differentbandwidth_spiral structure}. After applying the cut shown in Equation \ref{eq:vertical_cut}, we end up with 6704 stars. The distribution of the selected stars in the XY plane is displayed in the left panel of Figure \ref{fig:spiral_structure}, with individual stars represented by black points. 

Following the approach adopted in our previous work \citep[][hereafter P21]{Poggio:2021}, we map the stellar \emph{overdensity} $\Delta_{\Sigma}$, defined as
\begin{equation}
    \Delta_{\Sigma} (X,Y) = \frac{ \Sigma (X,Y) \, -  \langle \, \Sigma (X,Y) \, \rangle }{\langle \, \Sigma (X,Y) \, \rangle } = \frac{ \Sigma (X,Y) }{\langle \, \Sigma (X,Y) \, \rangle } - 1
    \quad,
    \label{Eq:overdensity}
\end{equation}
where $\Sigma(X,Y)$ is the local surface density at the position $(X,Y)$ in the Galactic plane, and $\langle \, \Sigma (X,Y) \, \rangle$ is the mean surface density. Both $\Sigma(X,Y)$ and $\langle \, \Sigma (X,Y) \, \rangle$ are calculated using a bivariate kernel density estimator, but using two different bandwidths, i.e. 0.6 kpc and 2 kpc, respectively, for the local and mean surface density. We refer the reader to Appendix B1 of \citetalias{Poggio:2021} for details on the calculation of $\Sigma(X,Y)$ and $\langle \, \Sigma (X,Y) \, \rangle$, which we do not repeat here for brevity. 

The right panel of Figure \ref{fig:spiral_structure} shows the regions of positive overdensity, $\Delta_{\Sigma}>0$, which correspond to areas typically more populated than the average density of stars at a given position in the Galactic disc. Consistent with expectations for young stellar populations, our dataset does not display a smooth distribution; instead, we clearly identify substructures that presumably correspond to segments of the nearest Galactic spiral arms. The stars preferentially lie along arc-like features and clumps, that are barely discernible in the raw spatial distribution (left panel of Figure \ref{fig:spiral_structure}), but become clearly visible once the over-dense regions of the Galactic disc are mapped. The sensitivity of the obtained map to different choices of adopted bandiwtdhs is discussed in Appendix \ref{appendix:impact_verticalcuts_differentbandwidth_spiral structure}.

The overdensity contours shown in the right panel of Figure \ref{fig:spiral_structure} exhibit clumpy but coherent structures, that extend over the volume covered by our dataset. Some of the observed clumpiness may be intrinsic to the Milky Way, while some may arise from observational effects. For instance, because Gaia data are in the optical band, they are expected to be significantly affected by interstellar extinction. Notably, extinction can cause a lack of stars along certain lines-of-sights, producing 'shadow cones' in the stellar distribution. Moreover, the selection function of the dataset can also have an impact, limiting our view in specific regions of the sky. To partially mitigate this effect, it is useful to compare the overdensity map derived in this work with maps based on other datasets that have different selection functions, thereby allowing the samples to be used in a complementary manner.

\begin{figure*}
    \centering
    \includegraphics[width=1\linewidth]{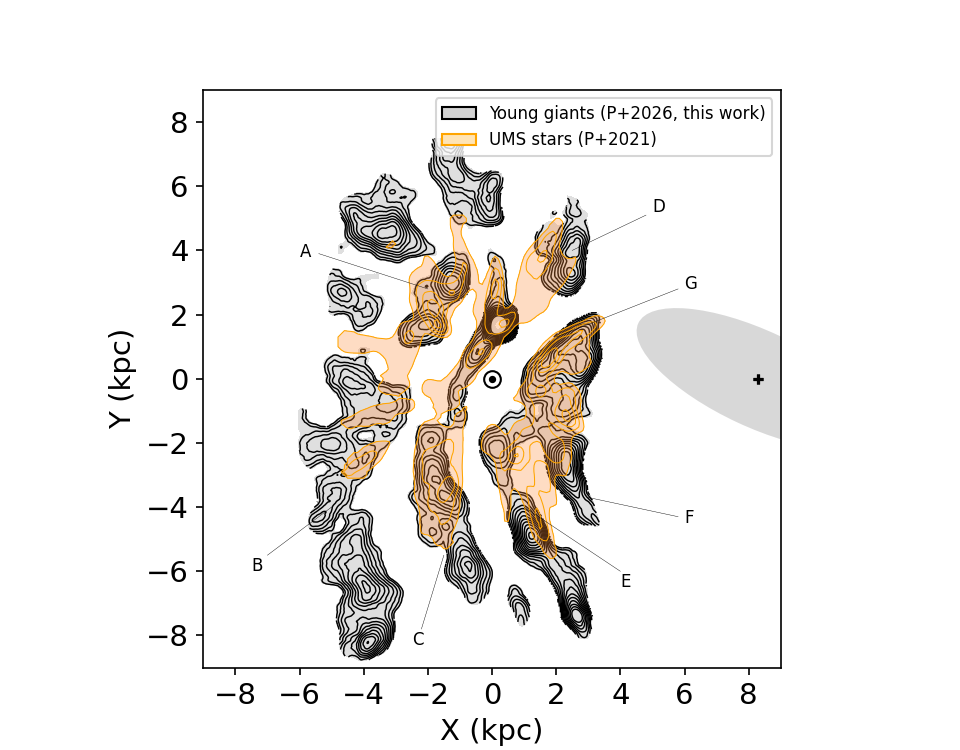}
    \caption{Same as the right panel of Figure \ref{fig:spiral_structure}, but now using transparent grey-shaded contours and black lines, overplotted on the overdensity contours obtained with Upper Main Sequence (UMS) stars from \citet{Poggio:2021}, for comparison. The Solar System is located in $(X,Y)=( 0\,  \rm{kpc}, 0 \, \rm{kpc})$. The black cross shows the position of the Galactic center, while the grey-schaded ellipse shows the relative orientation of the central bar of the Milky Way, as described in the text.}
    \label{fig:spiral_structure_YG_UMS}
\end{figure*}

Figure \ref{fig:spiral_structure_YG_UMS} shows a comparison between the overdensity contours obtained with our sample of young giant stars (the same as the right panel of Figure \ref{fig:spiral_structure}, but now using grey transparent contours) and those derived with Upper Main Sequence (UMS) stars in \citetalias{Poggio:2021}, shown in orange. As we can see, the contours obtained with the two samples appear to be in broad agreement in the regions where the two samples overlap. As discussed in \citetalias{Poggio:2021}, the UMS red contours exhibit three inclined stripes, corresponding (from left to right, respectively) to the Perseus arm, the local (Orion) arm closest to the Sun, and an inner band to the Sagittarius-Carina and Scutum arms. 

To facilitate the description of the young giants map, we associate labels to different regions, as indicated in Figure \ref{fig:spiral_structure_YG_UMS}. The feature labelled as `A' corresponds to the Cassiopeia region \citep[see for comparison Fig. 11 in][]{Zari:2021}, and it coincides with an analogous overdensity based on the UMS contours. This region has historically been identified as part of the Perseus arm. The feature labelled as `B' is quite extended, and spans from approximately $(X,Y)\sim(-3 \, \rm{ kpc}, 0 \, \rm{ kpc})$ to $(X,Y)\sim(-4 \, \rm{ kpc}, -8 \, \rm{ kpc})$. The outermost spiral arm segment traced by the UMS overdensity map, identified as a portion of the Perseus arm, not only coincides with feature `A', but also with the upper part of feature `B'. We note, however, that feature `A' and `B' traced by the young giants are clearly detached, whereas the UMS stars present a continuous distribution that bridges this gap. 

It is possible that the gap in the young giants is caused, at least in part, by lower completeness of our catalog in that region of the sky. For instance, we identified two prominent holes over a few degrees in the XGBoost distribution at low galactic latitudes,  one at $l\sim 154\degree - 148\degree, b\sim-3\degree - 5\degree$, and another one at $l\sim 275\degree - 278\degree, b\sim0\degree - 5\degree$. Those holes most likely originate from failed processing when XGBoost was constructed, e.g., due to timeouts during the crossmatch between XP spectra and other Gaia data (such as G, parallax, colour, etc.). The first hole does lie between features `A' and `B'; however, the observed gap is significantly larger than the hole itself, so cannot be fully explained by this issue.   


In general, within 3-4 kpc of the Sun, the UMS catalog is expected to be more complete than the young giant sample, since the selection function of the XGBoost catalog is less homogeneous. On the other hand, thanks to their large intrinsic brightness, the young giants can be mapped over larger portions of the Galactic disc compared to the UMS stars, from 2 to 5 kpc, depending on the considered direction. In the case of the Perseus arm, the young giant stars extend the segment traced by the UMS by about 5 kpc in length, from approximately $(X,Y)\sim(-4 \, \rm{ kpc}, -3 \, \rm{ kpc})$ to $(X,Y)\sim(-4 \, \rm{ kpc}, -8 \, \rm{ kpc})$. Only by overplotting both the UMS and young giant samples we can trace a continuous Perseus arm, which appears as a clumpy yet spatially coherent structure extending over more than 12 kpc. In contrast to the portion of the Perseus arm traced by the UMS sample, which could be approximated as a short, nearly straight segment, the view emerging from Figure \ref{fig:spiral_structure_YG_UMS} portrays the Perseus arm as an extended structure showing a gentle curvature, as expected for spiral arms on large scales.

Among the spiral arm segments traced by the UMS sample, the one nearest to the Sun is the Local (Orion) arm. This segment coincides for the most part with feature 'C' mapped with the young giants, extending from approximately $(X,Y)\sim(1 \, \rm{ kpc}, 2.5 \, \rm{ kpc})$ to $(X,Y)\sim(-1 \, \rm{ kpc}, -6.5 \, \rm{ kpc})$.  In the upper parts of the Local arm, there is a gap in the distribution of young giant stars, separating features C and D. It is unclear whether this gap is artificial 
or if it is an actual feature of the observed stars. 
In any case, similar to the gap between features 'A' and 'B', we see that this gap is bridged by the UMS sample. 
We also note that the overdensity of young giants extend the Local (Orion) arm mapped by the UMS sample by about 2 kpc from approximately $(X,Y)\sim(-1.5 \, \rm{ kpc}, -5 \, \rm{ kpc})$ to $(X,Y)\sim(0 \, \rm{ kpc}, -7 \, \rm{ kpc})$. 
When the UMS and young giant samples are overplotted, the Local arm appears as a coherent yet clumpy structure, extending over more than 10 kpc.

The inner regions of the map are more difficult to interpret. Feature 'E' traced by the young giants agrees well with the segment of the Sagittarius-Carina mapped with UMS stars in the region where they overlap. However, it is more extended in the lower parts of the plot by at least 2-3 kpc. Interestingly, the distribution of the young giant sample exhibits another feature, labelled as 'F' in Figure \ref{fig:spiral_structure_YG_UMS}, which appears to be a different arm segment, presumably corresponding to the Scutum arm. Thus, contrary to the UMS contours, the distribution of young giants presents two separated structures 
clearly detached in the lower parts of the plot ($Y < -2$ kpc), which seem to converge in the upper part of the plot. 
This is perhaps more clearly visible in the right panel of Figure \ref{fig:spiral_structure}, which is color-coded by overdensity. Meanwhile, feature `G' coincides with the high pitch angle feature discussed in \cite{Kuhn:2021}, which was mapped using Young Stellar Objects (YSOs) from \cite{Kuhn:2021_YSOcatalog} and masers from \citet{Reid:2019}.
In the young giants this nearly linear feature is seen to extend even further inward, showing a total length of at least 2 kpc. (See figure in Appendix \ref{appendix:comparison_with_Kuhn}.) Indeed, the clear gap between features `G' and `D' gives the impression that the linear feature `G' is the Sagittarius arm tangent at positive galactic longitudes, resulting from an abrupt change in pitch angle of the arm. 

Finally we point out three disconnected features in the upper left corner of Figure \ref{fig:spiral_structure_YG_UMS} that lie beyond the Cassiopeia region (Feature `A') mentioned above. Here our young giants may be showing us pieces of the Outer Arm beyond the Perseus arm. 

To provide the reader with a more complete large-scale view of the Milky Way, the grey-shaded ellipse in Figure \ref{fig:spiral_structure_YG_UMS} shows the orientation of the Milky Way's central bar. Given that there is currently no consensus on the structural parameters of the Galactic bar \citep[see the reviews by][ and references therein]{BlandHawthorn:2016, HuntVasilev:2025}, here we rely on some representative values, adopting a major axis of $\sim$4 kpc and a viewing angle of $\alpha_b \sim 25^{\degree}$  \citep[][]{HuntVasilev:2025}, to show its orientation with respect to the spiral structure mapped in this work.

\begin{figure*}
    \centering
    \includegraphics[width=\linewidth]{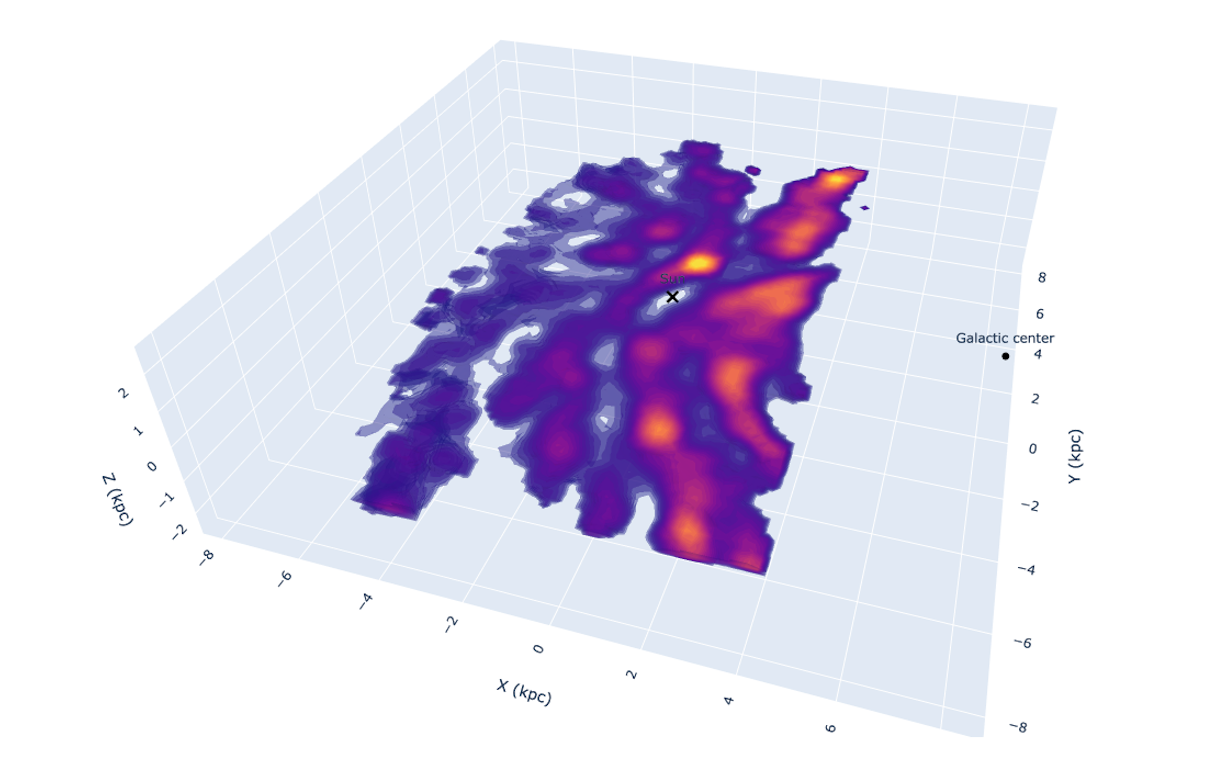}
    \includegraphics[width=\linewidth]{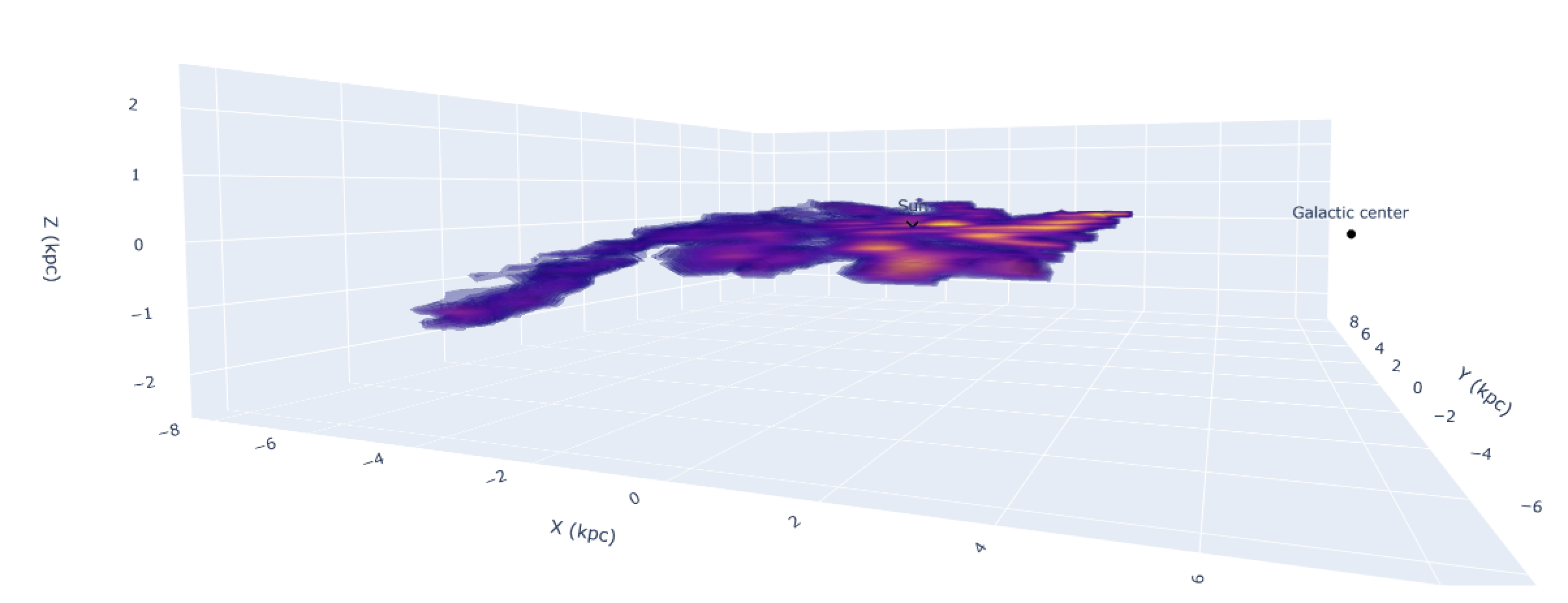}
    \caption{Three-dimensional view of the overdense regions as mapped with the young giant sample (see text for more details). The two panels show two different examples of viewing angles. The corresponding interactive three-dimensional plot will be made available.}
    \label{fig:3D}
\end{figure*}

\begin{figure*}
    \centering
    \includegraphics[width=\linewidth]{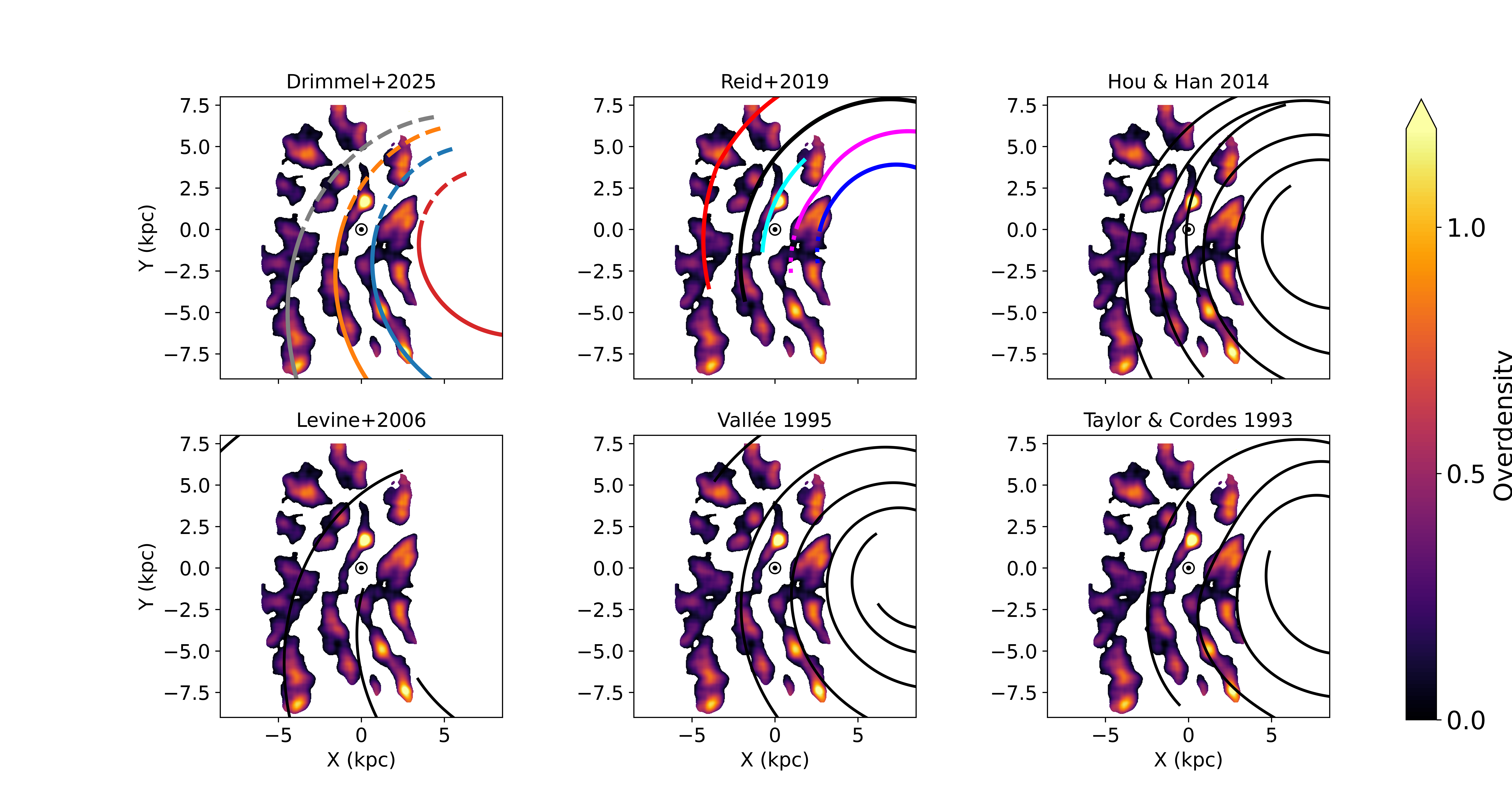}
    \caption{Comparison between the spiral structure mapped in this work and other works available in literature \citep{Drimmel:2025, Reid:2019, Hou:2014, Levine:2006, Taylor:1993, Valle:1995}, as indicated by the title of each panel. Models are plotted using the Python library SpiralMap \citep{Prusty:2025}.}
    \label{fig:spiral_structure_comparison_models}
\end{figure*}

\subsection{Three-dimensional view of the spiral structure}\label{sec:3Dspiral}

To obtain a more comprehensive view of the Galactic disc, in this subsection we extend the two-dimensional approach adopted in Section \ref{sec:spiral} to three dimensions. A comparable methodology, but centered on metallicity, was employed by \cite{MartinezMedina:2025}, who constructed a three-dimensional map of the metallicity excess within 3–4 kpc of the Sun and used it as a primary diagnostic to investigate the morphology and chemical enrichment of the spiral arms. In this contribution, we simply focus on the three-dimensional distribution of stars in our sample. Equation \ref{Eq:overdensity} therefore becomes:
\begin{equation}
    \Delta_{\rho} (X,Y,Z) = \frac{ \rho (X,Y,Z) \, -  \langle \, \rho (X,Y,Z) \, \rangle }{\langle \, \rho (X,Y,Z) \, \rangle } = \frac{ \rho (X,Y,Z) }{\langle \, \rho (X,Y,Z) \, \rangle } - 1
    \quad,
    \label{Eq:overdensity_3D}
\end{equation}
where $\rho(X,Y,Z)$ is the three-dimensional local stellar density and $\langle \, \rho (X,Y,Z) \, \rangle$ is the three-dimensional mean stellar density, which are computed as described below. For a given position $(X,Y,Z)$, we compute the local density $\rho(X,Y,Z)$ through a multivariate Kernel Density Estimator \citep[following Eq. 6.11 in][]{Feigelson:2012} in three dimensions, starting from the $(x_i, y_i, z_i)$-coordinates of the N stars in our sample, where $i=1,...., N$:
\begin{equation}
 \rho(X,Y,Z)  = \frac{1}{N \, h_X h_Y h_Z } \sum_{i=1}^{N} \Bigg[  K\bigg( \frac{X-x_i}{h_X} \bigg) \,  K\bigg(\frac{Y-y_i}{h_Y}\bigg)  \,  K\bigg(\frac{Z-z_i}{h_Z}\bigg) \Bigg] \quad,
 \label{loc_dens}
\end{equation}
where $K$ is the kernel function and $h_X, h_Y, h_Z$ are, respectively, the kernel bandwidths along the X, Y and Z directions. We adopted an Epanechnikov kernel function
\begin{equation}
 K\bigg( \frac{X-x_i}{h_X} \bigg) = \frac{3}{4} \, \bigg( 1 - \bigg( \frac{X-x_i}{h_X} \bigg)^2 \bigg)
\end{equation}
for $ |(X-x_i)/h_X|<1$, and zero outside (and similarly for the Y- and Z-coordinate). Given the geometry of the problem under consideration, here we adopt the same bandwidth for the X and Y coordinates, but use a smaller bandwidth along the Z coordinate to account for the thinness of the Galactic disc. To estimate the mean density $\langle \, \rho (X,Y,Z) \, \rangle$, we use the same approach adopted for the local density (Equation \ref{loc_dens}), but choose a larger bandwidth. Following Equation \ref{Eq:overdensity_3D}, the local and mean densities are combined, to derive the three-dimensional stellar overdensity at a given position (X,Y,Z). Figure \ref{fig:3D} shows the regions of positive overdensity ($\Delta_{\rho} (X,Y,Z)>0$) using bandwidths of 0.6 and 2 kpc, respectively, for the local and mean density along the X and Y direction, and 0.1 and 0.5 kpc along the Z direction. Other values of bandwidths were tested, but the corresponding maps are not shown here for brevity. Additionally, we also tested other kernel functional forms, such as the triangle and Gaussian kernel \citep[see for example][]{Feigelson:2012}, to ensure that the choice of kernel does not significantly affect the resulting maps. 

The upper and lower panels of Figure \ref{fig:3D} show two different viewing angles of the stellar overdensity in three-dimensions. (An interactive plot is available to examine the map at different viewing angles and with zoom-in/out options.) As expected, the map shown in Figure \ref{fig:3D} exhibits an in-plane distribution of stars consistent with the bi-dimensional map in the XY plane shown in Figure \ref{fig:spiral_structure} (right panel) and Figure \ref{fig:spiral_structure_YG_UMS}. 
However, Figure \ref{fig:3D} visually provides additional information on the vertical coordinate, showing evidence of a vertically warped spiral structure. This is mostly evident in the outer parts of the disc, where the Galactic warp is stronger. It is therefore the Perseus arm, together with a few other features in the outer disc potentially associated with the Outer arm, that exhibit the strongest vertical deviations from the Z=0 plane. From the viewing angle of the lower panel in Figure \ref{fig:3D}, one can clearly see that the visible portion of the Perseus arm (corresponding to Feature B in Figure \ref{fig:spiral_structure_YG_UMS}) is vertically warped downwards ($Z<0$), and clearly detached from the nearest segment of the Local arm (labelled as Feature C in Figure \ref{fig:spiral_structure_YG_UMS}).

\section{Discussion} \label{sec:discussion}

In Figure \ref{fig:spiral_structure_comparison_models} we overplot our overdensity map of the young giants with some of the spiral models currently found in the literature. As we can see, none of them agree perfectly. The Sagittarius, Local Arm and Perseus arm as traced by the Classical Cepheids \citep{Drimmel:2025} overlap quite well with young-giant overdensity, but not the innermost Scutum arm, which is most affected by reddening, possibly causing bias in the photometric distances of the Cepheids. Alternatively the Cepheids inside the Sagittarius arm are tracing another arm at even smaller radii. Only the model of \citet{Reid:2019} places this innermost arm traced by the young giants at the same location. The Reid model also traces an Outer Arm, which coincides with the three outer-most overdensity "blobs" in the third quadrant. However, compared to the model from \citet{Reid:2019}, the contours obtained in this work suggest a more open geometry (i.e. larger pitch angle) for the Local and Perseus arms, in agreement with the model of \citet{Drimmel:2025}. 

The spiral arm model of \citet{Levine:2006}, based on the kinematic distances of HI gas, successfully traces the Perseus arm (feature 'A' and 'B' in Figure \ref{fig:spiral_structure_YG_UMS}), but is less successful in tracing the other features mapped by the young giants. The other spiral arm models shown in Figure \ref{fig:spiral_structure_comparison_models} \citep{Taylor:1993, Valle:1995, Hou:2014} only trace reasonably well the Sagittarius-Carina arm. These models usually also trace the lower part of the feature labelled as 'C' in Figure \ref{fig:spiral_structure_YG_UMS}, which they typically assign to the Perseus arm; on the contrary, our maps suggest that it would be more natural to consider this feature as part of the Local arm.  Meanwhile only the \citet{Drimmel:2025} model based on the Cepheids trace the Local Arm over its full extent. Historically this feature was thought to be a small spur or bridge between the Sagittarius and Perseus arm. Indeed both \citet{Reid:2019} and \citet{Hou:2014} only model it over a short extent. However, already in \citetalias{Poggio:2021} using the UMS sample, the Local Arm was traced as a continuous feature over a length of 8 kpc. Here we see that it is at least 10 kpc in length, while both \citet{Reid:2019} and \citet{Hou:2014} assign the lower segment of the Local Arm in the third quadrant to the Perseus Arm.

\begin{figure}
    \centering
    \includegraphics[width=\linewidth]{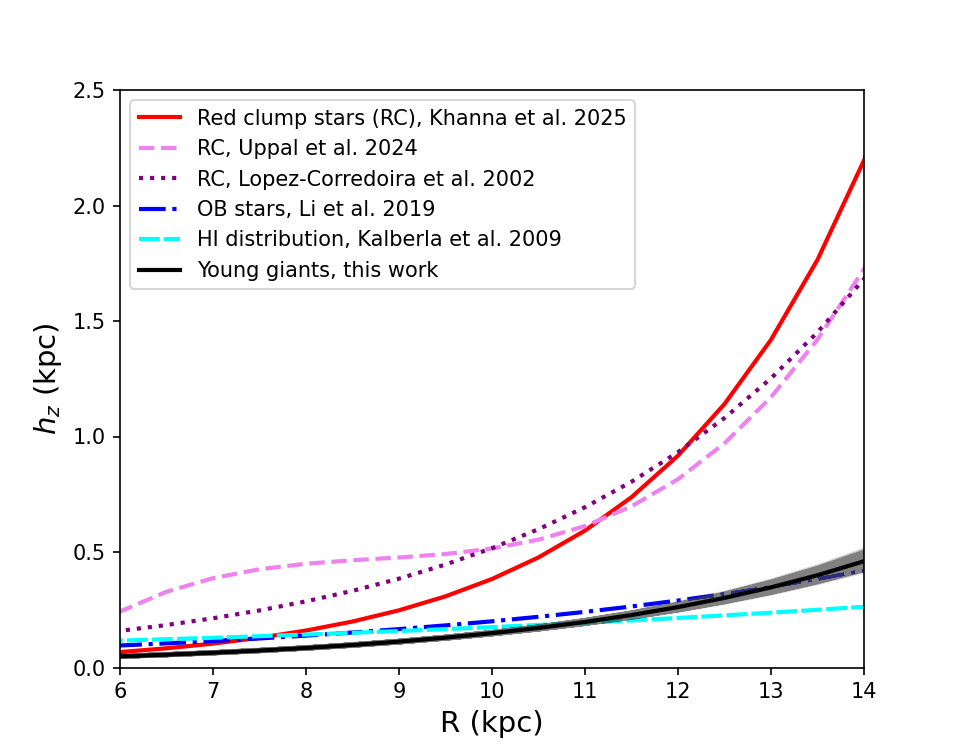}
    \caption{Comparison between the flare parametrization obtained in this work (black line/grey shaded area) and other models available in literature for different tracers.}
    \label{fig:comparison_flare}
\end{figure}

\begin{figure*}
    \centering
    \includegraphics[width=\linewidth]{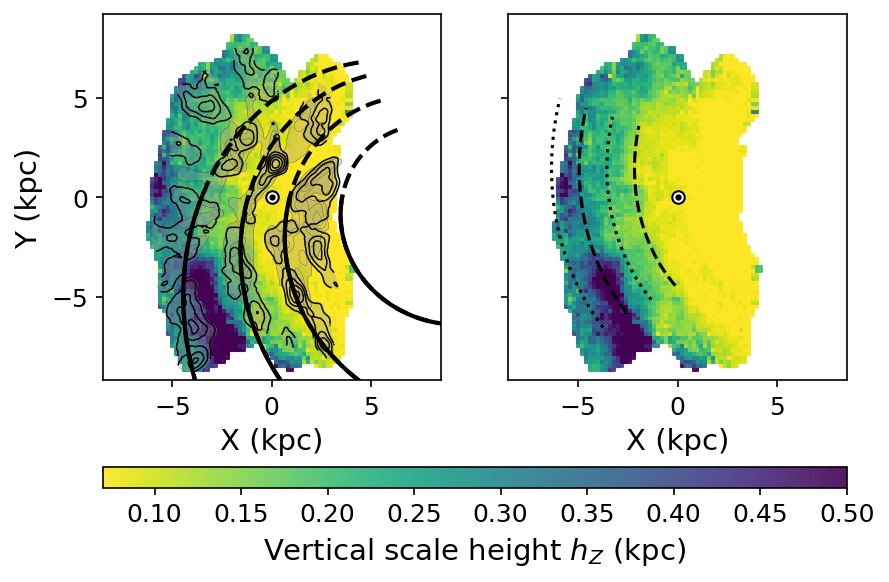}
    \caption{\emph{Left panel:} Same as Figure \ref{fig:hz_in_xyplane_overlapping_cells}, but now compared to the overdensity contours from the right panel of Figure \ref{fig:spiral_structure} (black contours, greay shaded areas). The overdensity contours from Upper Main Sequence stars in \citet{Poggio:2021} (thin grey contours, grey shaded areas) and the spiral arms model from \citet{Drimmel:2025} (black lines) based on Cepheids are also shown for comparison. \emph{Right panel:} Same as Figure \ref{fig:hz_in_xyplane_overlapping_cells}, but now compared to the position of the vertical wave mapped in \citetalias{Poggio:2025}. The meaning of the dashed and dotted lines is explained in the text. }
    \label{fig:comparison_hz_spiral_arms_great_wave}
\end{figure*}

In addition to the spiral structure, it is also interesting to discuss and compare the flare parametrization obtained in this work with other models based on different tracers available in the literature. 
In this work, we have exploited a high fidelity sample of young giants, with typical stellar ages of approximately less than 100-200 Myr. Our sample has the advantage of being both numerous, as well as covering a large span of the disc ($5<R$ [kpc]$\, < 14$). We have parameterized the Galactic flare such that the scale-height is exponential in $R$, and the best fitting profile is shown in \autoref{fig:comparison_flare} as a solid black curve characterised by $h_{z0} \sim$77 pc and a flare scale-length of $h_{fl} \sim$3.5 kpc. This can be compared with the blue dash-dotted curve, which is the flare profile of OB stars from Gaia DR2 obtained by \cite{Li:2019}. Since our sample consists of very young stars, we have also included a comparison with the flare profile inferred from the Milky Way's HI distribution by \cite{2009ARA&A..47...27K}, shown as a cyan dashed curve. In both cases we find excellent agreement with other studies of the younger components of the disc in both stars and gas.

The Galactic flare has also been studied using more evolved stellar populations such as the red clump (RC), mainly due to their large sample sizes ($\sim$ millions) as well as being a standard-candle, allowing reliable distance estimation. 
With the RC sample we can trace the properties of the older disc, as they have a broad range in ages between 1-5 Gyr 
\citep{girardi_rc16,warfield2024_rc}. In \autoref{fig:comparison_flare}, we show the flare profile from three different studies using the RC stars \citep{LopezCorredoira:2002, Uppal:2024, Khanna:2025}. 
All of these studies find a much stronger flare in the RC population than in the young giants, as one would expect. However, none of these models of the flare in the RC population took account of the warp, so it is likely that the strength of the flare is overestimated.


Figure \ref{fig:xy_map_hz_results_fits_cells} (right panel) and Figure \ref{fig:hz_in_xyplane_overlapping_cells} show that the variation of the scale height is not axisymmetric. In the portion of the outer disc covered by our sample, the disc  is thicker in the 
third quadrant ($X,Y<0$) 
than in the second quadrant ($X<0,Y>0$) 
between 12 and 14 kpc in Galactocentric radius. 
In particular, there is a portion of the disc in the third quadrant with a significantly higher scale height, and this region roughly corresponds to the inter-arm region between the Local arm and Perseus arm, as indicated in 
the left panel of Figure \ref{fig:comparison_hz_spiral_arms_great_wave}, showing the comparison between the disc thickness $h_Z$ 
and the density contours constructed from our young giant sample 
(black contours) and the UMS sample of P21 (grey shaded areas). 
We also show, for comparison, 
the spiral arm model from \citet{Drimmel:2025} based on classical Cepheids (black lines). 
It remains unclear whether this anti-correlation of the scale height with stellar density provides any insight into the origin of this observed feature. For instance, it is possible that the relative fraction of contaminants (which in general is expected to be very small, as discussed in Section \ref{sec:flare}) is larger in the interarm regions compared to the areas along the spiral arms, leading to larger values of vertical thickness. However, we checked the relative fraction of contaminants as mapped by the $N_{floor}$ parameter in each XY-cell (see Section \ref{sec:flare}) and found that this area of large $h_Z$ 
exhibits low values of $N_{floor}$, both in absolute terms and with respect to the total number of stars per cell, indicating that this region is not expected to be dominated by contaminants. 

To consider an alternative (though not necessarily mutually exclusive) explanation, we also explored the possible connection with known vertical distortions of the disc. Specifically, the right panel of Figure \ref{fig:comparison_hz_spiral_arms_great_wave} shows the comparison between the position of the large-scale vertical corrugation mapped in \citetalias{Poggio:2025}. The dashed and dotted lines are the same as the ones shown in the right panels of Fig. 12 and 13 of \citetalias{Poggio:2025}. We refer the reader to that paper for a detailed explanation of the meaning of those curves, but we provide here a very brief summary for the sake of clarity. In \citetalias{Poggio:2025}, we fitted for a classical $m=1$ warp vertical distortion and then explored the fit residuals $\Delta_Z$. We found a region of systematically positive residuals, indicating a portion of disc systematically shifted upwards with respect to the classical warp, roughly delimited by the two black dashed lines. To the left of this region, the vertical residuals were found to be systematically negative ($\Delta_Z < 0$), consistent with a nearly radial large-scale vertical ripple or corrugation. The leftmost dashed line marks the transition between positive and negative residuals, therefore showing a region where the vertical corrugation changes its sign. We note that this line partially overlaps with the regions of large $h_Z$ shown by our map. The two dotted lines are related to the vertical velocities, and roughly delimit the region of positive residuals in the vertical velocities $\Delta_{V_Z} > 0$, and are radially shifted with respect to the positive $\Delta_Z$, suggesting a wave-like behaviour, as discussed in \citetalias{Poggio:2025}. While it is possible that the observed azimuthal asymmetry in $h_Z$ mapped in this work is related to the wave mapped in \citetalias{Poggio:2025}, no definitive conclusions can be drawn at this stage.

\section{Conclusion} \label{sec:conclusion}

In this contribution, we have analyzed the three-dimensional distribution of a sample of young giant stars selected from Gaia DR3 data that cover a significant fraction of the Galactic disc. Our main results can be summarized as follows:

\begin{itemize}

    \item \emph{Young giant stars in the Galactic disc clearly exhibit an exponential flare.} Taking account of the selection function of our sample and correcting for the Galactic warp, we show that the flare
    can be modelled as follows: $h_z(R) = h_{Z0} \exp{(R-R_{\odot})/h_{fl}}$, where the flare's characteristic radial scale length $h_{fl} =  3.5 \pm 0.3 \, \rm{kpc}$ and the vertical scale height of the young Galactic disc at the Sun's position $h_{Z \odot} = 77 \pm 4 $ pc. The resulting flare is fairly consistent with the measured flare of a sample of OB stars by \citet{Li:2019}, as well as that seen in the HI disk \citep{2009ARA&A..47...27K}.  
    
    \item \emph{
    In addition to the exponential flare, there is evidence of azimuthal variations in the thickness of the young Galactic disk.}  These variations are seen at Galactocentric radii beyond $\sim$12 kpc, and are most evident in the third Galactic quadrant beyond 14 kpc.
    The origin of the observed azimuthal asymmetry is unclear. Several possible causes are discussed, including the presence of the spiral structure in our sample and/or the influence of vertical distortions in the Galactic disc, such as the warp and vertical corrugations (see Section \ref{sec:discussion}).
    
    \item \emph{The spiral structure of the Galaxy is mapped out to $\sim$ 8 kpc.} The in-plane distribution of the sources traces coherent spiral arm segments, extending previous maps based on upper main-sequence (UMS) and OB stars \citep{Poggio:2021,Zari:2021,GaiaCollDrimmel:2023} by 2 to 4 kpc, depending on the considered direction. The Perseus Arm is now clearly traced over 12 kpc with a spiral pitch angle of roughly 20 degrees, in agreement with the mapping of this arm with Cepheids \citep{Drimmel:2025} and HI \citep{Levine:2006}. The Local/Orion arm is shown to extend over a length of at least 10 kpc, about 2 kpc further than previously mapped. Meanwhile in the inner Galaxy we identify a new arm segment likely associated with the Scutum Arm, clearly detached from the Sagittarius–Carina Arm in the fourth Galactic quadrant. The map presented in Figure \ref{fig:spiral_structure} and \ref{fig:spiral_structure_YG_UMS} will be made available.

    \item \emph{We present a new three-dimensional map of the young Galactic disc.} Our dataset offers a three-dimensional view of our Galactic disc, simultaneously showing evidence of the spiral structure, the warp (discussed in \citetalias{Poggio:2025}) and the flare. While these elements are often studied separately in the literature, this study aims to give a comprehensive view of the Galactic disc, which is crucial to properly determine its geometric properties. Indeed, we show that neglecting the warp would lead to significant over-estimate of the scale height. An interactive three dimensional map of our dataset will be made available.
\end{itemize}


Collectively, the above mentioned results improve our knowledge of the structure of the Milky Way's thin disc on large scales, as mapped by young populations, especially in the third Galactic quadrant. Future datasets with improved coverage of the Galactic disc, such as those expected from Gaia Data Release 4 and 5, in synergy with complementary surveys, potentially combined with full three-dimensional stellar kinematics, will enable further extensions of these maps and provide tighter observational constraints on the structure and evolution of the Milky Way.

\begin{acknowledgments}
SK, RD and EP were supported in part by the Italian Space Agency (ASI) through contract ASI-INAF 2025-10-HH.0 to the National Institute for Astrophysics (INAF). This work has made use of data from the European Space Agency (ESA) mission
\gaia\ (\url{https://www.cosmos.esa.int/gaia}), processed by the \gaia\
Data Processing and Analysis Consortium (DPAC,
\url{https://www.cosmos.esa.int/web/gaia/dpac/consortium}). Funding for the DPAC
has been provided by national institutions, in particular the institutions
participating in the \gaia\ Multilateral Agreement.

\end{acknowledgments}

\appendix

\section{Individual likelihood approach}\label{subsec:individual_likelihood}

An alternative approach to the methodology outlined in Section \ref{subsec:Zprimebinning} is to compute the probability to find each individual star of a given XY-cell at the vertical position $Z^{\prime}$, that is
\begin{equation}
\label{eq:individual_likelihood}
\begin{split}
\lefteqn{P(Z^{\prime}|\sigma_Z, h_z, l,b,G) =} &  \\
&  \int^{+\infty}_{-\infty} C(l,b,G) \, p(Z^{\prime}|\sigma_Z, Z^{\prime}_{true}) \, p(Z^{\prime}_{true}|h_z) \, dZ^{\prime}_{true}\, ,
\end{split}
\end{equation}
where
\begin{equation}
\label{eq:individual_likelihood_modelgal}
p(Z^{\prime}_{true}|h_z) = \frac{1}{2 \, h_Z} e^{- \frac{ |Z^{\prime}_{true}|}{h_Z}  }
\end{equation}
is the model of the Galactic disc, and 
\begin{equation}
\label{eq:individual_likelihood_obs}
p(Z'|\sigma_Z', Z'_{true}) = \frac{1}{\sqrt{2 \pi} \, \sigma_Z'} e^{- \frac{(Z' - Z'_true)^2}{2 \, \sigma_Z'^2}  }
\end{equation}
is the model of the uncertainty, where $Z'_{true}$ is the true (unknown) position, $Z'$ is the observed position of the star, and $\sigma_Z' = (Z'_{84} -  Z'_{16})/2$, 
where $Z'_{84}$ and $Z'_{16}$ are, respectively, the $84^{th}$ and $16^{th}$ percentile of the $Z'$ distribution, computed for each star from the distance posterior distribution, and assuming no error on the observed galactic latitude $b$. The correction factor $C(l,b,G)$ in Equation \ref{eq:individual_likelihood} is computed similarly to the ${\mathcal{F}}_{i}$ factor described in Section \ref{subsec:Zprimebinning}, but now using the individual $(l,b,G)$ of the star. 

For each star, we perform the integral in Equation \ref{eq:individual_likelihood} over the true unknown $Z'_{true}$ via Monte Carlo integration. Once the individual likelihoods are computed for each star, we calculate the total likelihood as the product of all individual likelihoods (or, to be more precise, the sum of all log-likelihoods, as it is computationally more convenient to deal with logarithmic quantities). We adopt as best-fit result the value of $h_z$ that maximizes the total likelihood, and the $16^{th}$ and $84^{th}$ percentiles of the $h_z$ distribution as uncertainty on the estimated parameters. 

To estimate the goodness of this approach, we tested it using mock catalogs as described in Section \ref{subsec:mock_catalog}, but with less runs due to the longer computational time required using this methodology. We performed different tests at different distances and with different $N_{mock}$ stars, and we find that, if we do not include the $N_{floor}$ parameter (as in Equation \ref{eq:individual_likelihood_modelgal}), $h_{Z,TRUE}$ is usually recovered within 1 $\sigma$, and always within $2\sigma$. If the $N_{floor}$ parameter is included in the mock catalog and in the model (i.e. added to the right-hand side of Equation \ref{eq:individual_likelihood_modelgal}), we are able to recover the true value within $2-3\sigma$, but with a longer computational time (compared to the case without $N_{floor}$), given that the inference is now performed over 2 parameters. However, it is interesting to note that, if the $N_{floor}$ parameter is included in the mock catalog and not in the model, the result of the fit is biased by more than 5 $\sigma$. This might be due to the fact that this approach requires a very good knowledge of the true (but, \emph{a priori}, unknown) model of the Galaxy: if the assumed model is not correct, the approach tends to accomodate outliers, and the final result can be strongly biased. A similar situation is not happening with the $Z^{\prime}$-approach, which appears more robust to a possible presence of outliers, or a model that does not perfectly match the (\emph{a priori} unknown) underlying truth.

\section{Impact of different vertical cuts or adopted bandwidths on the spiral structure} \label{appendix:impact_verticalcuts_differentbandwidth_spiral structure}

\begin{figure*}
    \centering
    \includegraphics[width=0.33\linewidth]{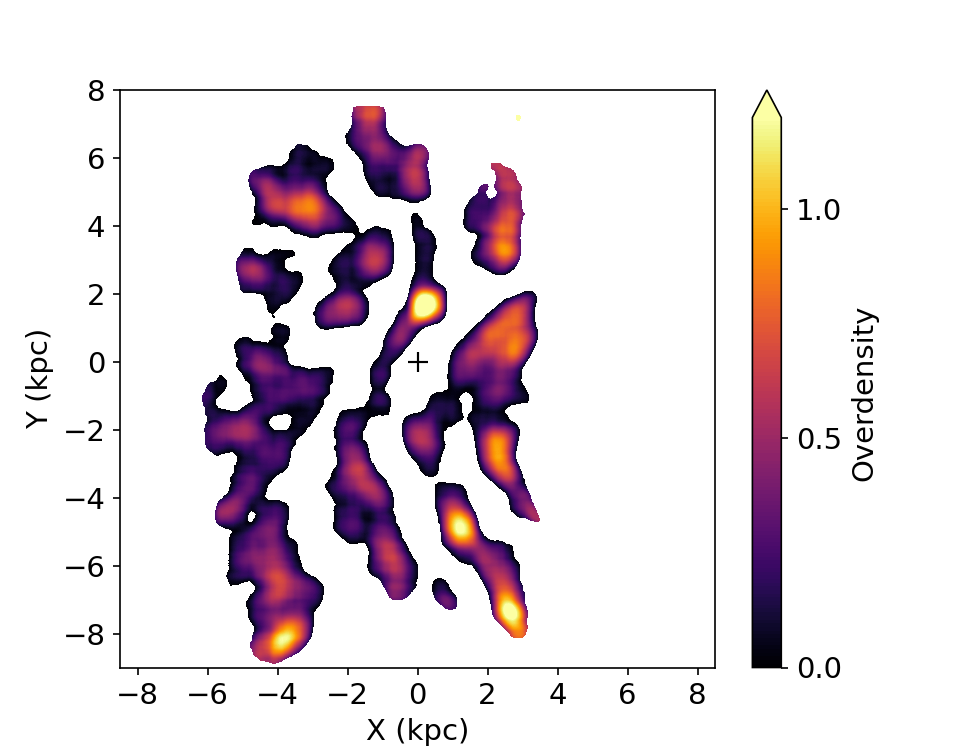}
    \includegraphics[width=0.33\linewidth]{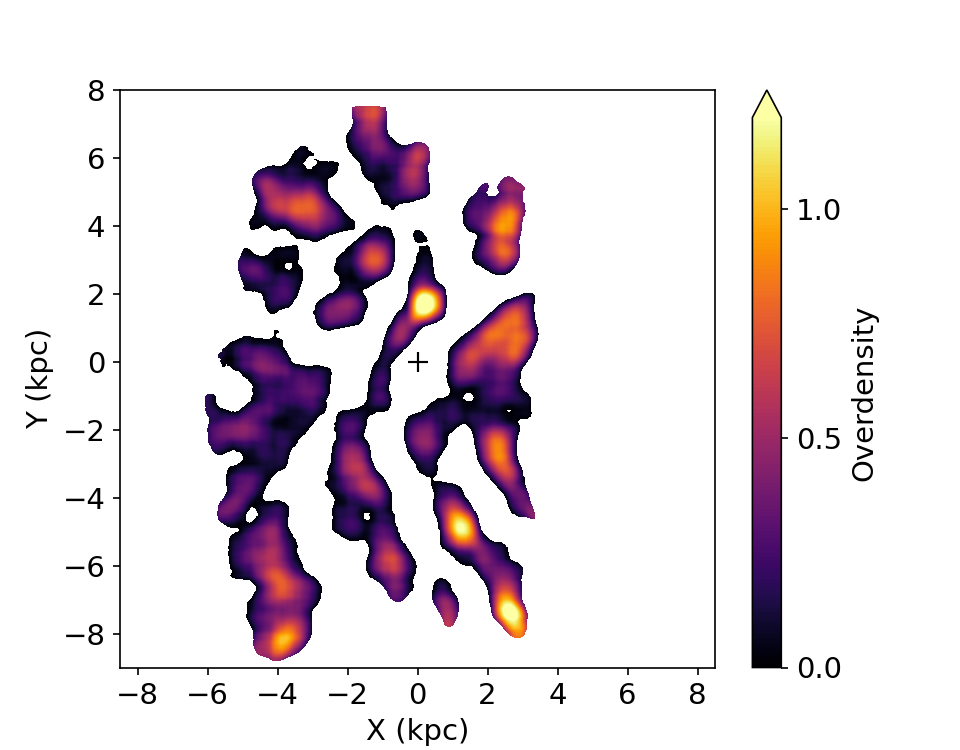}
    \includegraphics[width=0.33\linewidth]{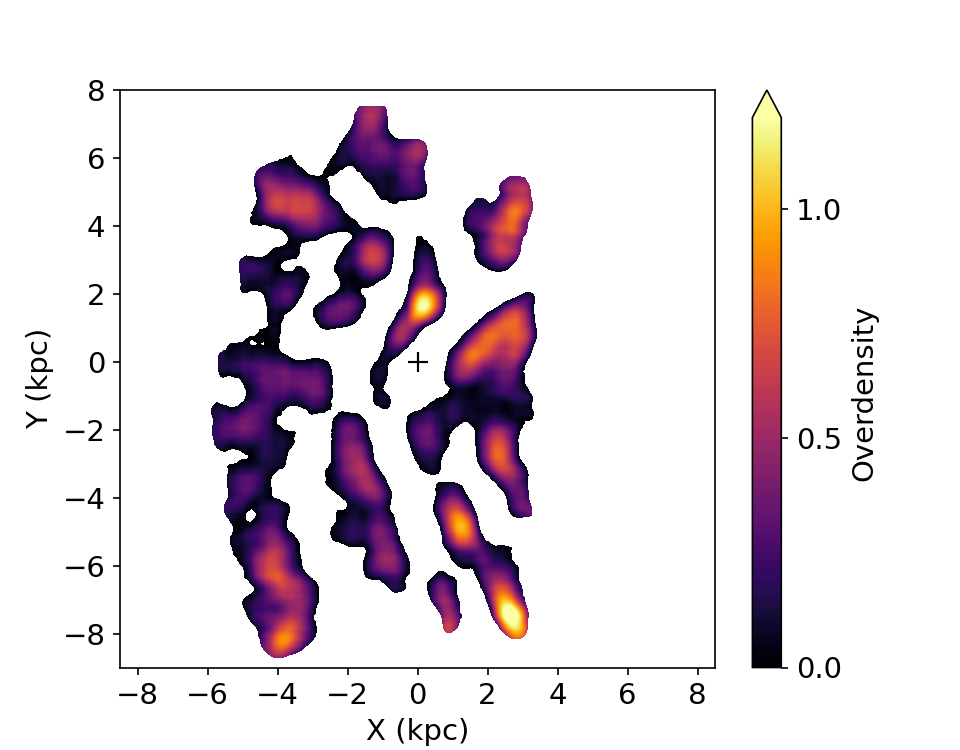}
    \caption{Same as the right panel of Figure \ref{fig:spiral_structure}, but now using different vertical cuts based on the following rescaling factors: 0.7 (left panel), 0.9 (middle panel) and 1.5 (right panel). }
    \label{fig:spiral_structure_different_cuts}
\end{figure*}

As discussed in Section \ref{sec:spiral}, the map of the spiral structure presented in this work is obtained by applying the vertical cut shown in Equation \ref{eq:vertical_cut}. This selection aims to isolate stars in the Galactic plane while accounting for vertical disc distortions (by rescaling the midplane using the median vertical coordinate $Z_{MED}$) and the variations in the vertical scale height $h_z$. However, Equation \ref{eq:vertical_cut} also includes a rescaling factor (taken as 0.8 in Section \ref{sec:spiral}), which is arbitrarily chosen. For this reason, here we test the impact of different possible choices. One can note that, if the rescaling factor is very large, the relative contribution from potential contaminants at large vertical heights can be higher, making the spiral structure in the Galactic plane less visible or noisier. Conversely, if the rescaling factor is too small, the sample size becomes insufficient, resulting in significantly noisier maps. While the value adopted in the main text was empirically chosen to balance these two effects, we present here the resulting maps for alternative choices to illustrate their sensitivity to this parameter. Figure \ref{fig:spiral_structure_different_cuts} shows, from left to right, the maps obtained rescaling factors of 0.7, 0.9 and 1.5, respectively.

\begin{figure*}
    \centering
    \includegraphics[width=0.33\linewidth]{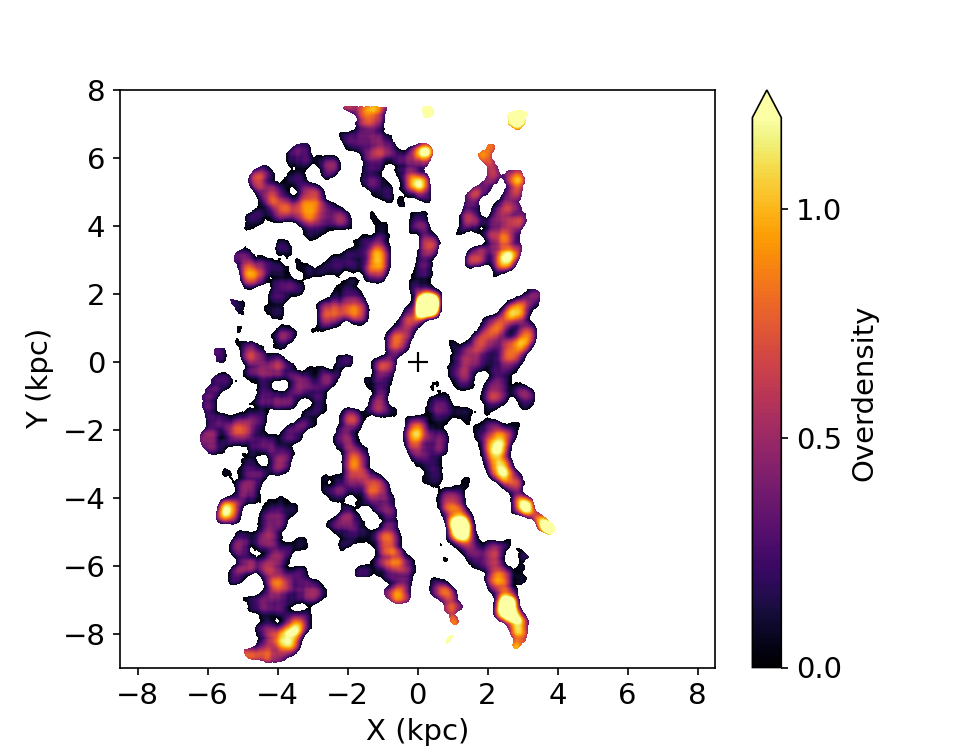}
    \includegraphics[width=0.33\linewidth]{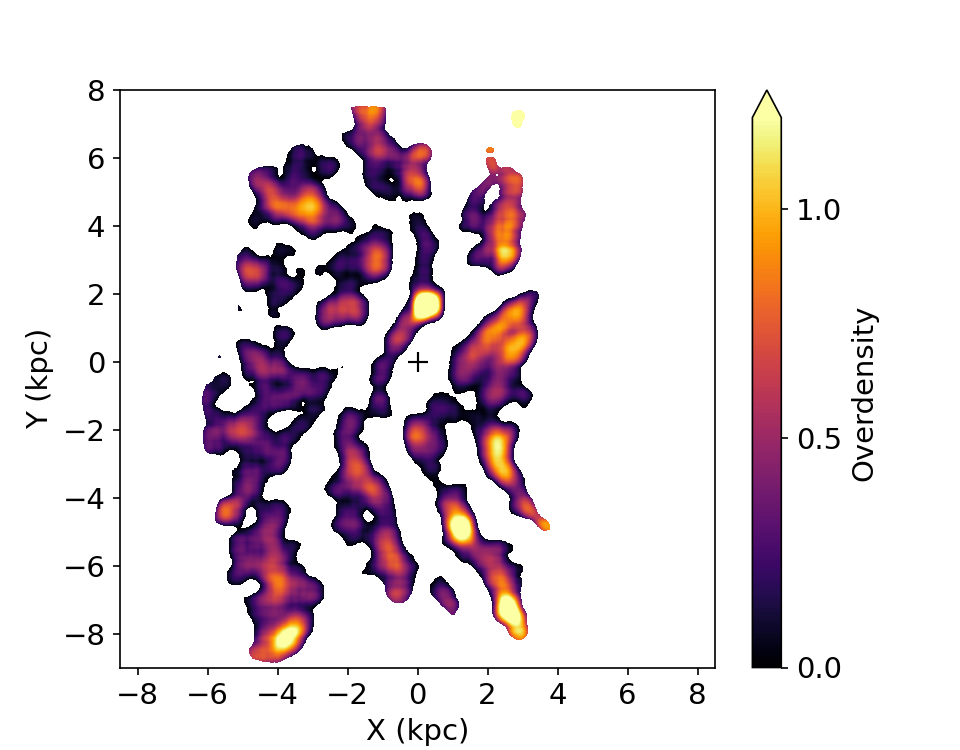}
    \includegraphics[width=0.33\linewidth]{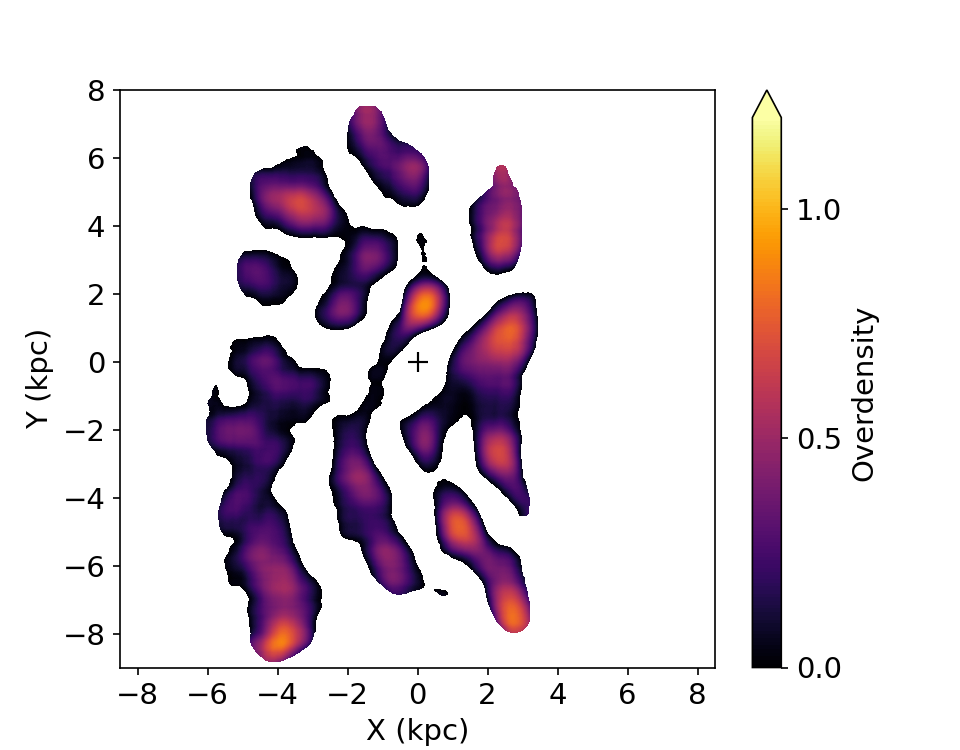}
    \caption{Same as the right panel of Figure \ref{fig:spiral_structure}, but now using different bandwidths: $w_{loc} = 0.4 \rm{\, kpc}$ and $w_{mean} = 1.6 \rm{\, kpc}$ (left panel), $w_{loc} = 0.5 \rm{\, kpc}$ and $w_{mean} = 2 \rm{\, kpc}$ (middle panel), and $w_{loc} = 0.8 \rm{\, kpc}$ and $w_{mean} = 2 \rm{\, kpc}$ (right panel).}
    \label{fig:spiral_structure_different_bandwidhts}
\end{figure*}

It is also important to discuss how different adopted bandwidths can influence the map(s) presented in this work. Of course, the choice of bandwidth is completely arbitrary, but some reasonable criteria could be adopted to select the values that best emphasize the presence of the spiral structure in our dataset. For instance, using a local bandwidth much larger than the typical size of the structure of interest (in this case, the typical width of the spiral arms of the Milky Way) would lead to an oversmoohting of the features. Conversely, adopting a narrower local bandwidth increases the risk of over-fitting to statistical fluctuations, yielding a noisier map. It should also be noted that the lower threshold for an acceptable bandwidth also depends on the typical number density of the sample, given that low-number statistics require a wider bandwidth to ensure reliable density estimates. Figure \ref{fig:spiral_structure_different_bandwidhts} shows the results obtained adopting, for the local and mean density, respectively, the bandwidths $w_{loc} = 0.4 \rm{\, kpc}$ and $w_{mean} = 1.6 \rm{\, kpc}$ (left panel), $w_{loc} = 0.5 \rm{\, kpc}$ and $w_{mean} = 2 \rm{\, kpc}$ (middle panel), and $w_{loc} = 0.8 \rm{\, kpc}$ and $w_{mean} = 2 \rm{\, kpc}$ (right panel). As already noted in previous works, the obtained maps are very sensitive to the choice of local bandwidths, while they exhibit a milder dependency on the bandwidth chosen for the mean density. For instance, as a test, we constructed a map using $w_{loc} = 0.4 \rm{\, kpc}$ and $w_{mean} = 2 \rm{\, kpc}$ (not shown here). The resulting map is very similar to the left panel of Figure \ref{fig:spiral_structure_different_bandwidhts}, which shares the same $w_{loc}$, but employs a different $w_{mean}$.

\section{Comparison with Kuhn et al. (2021)} \label{appendix:comparison_with_Kuhn}

\begin{figure}
    \centering
    \includegraphics[width=0.75\linewidth]{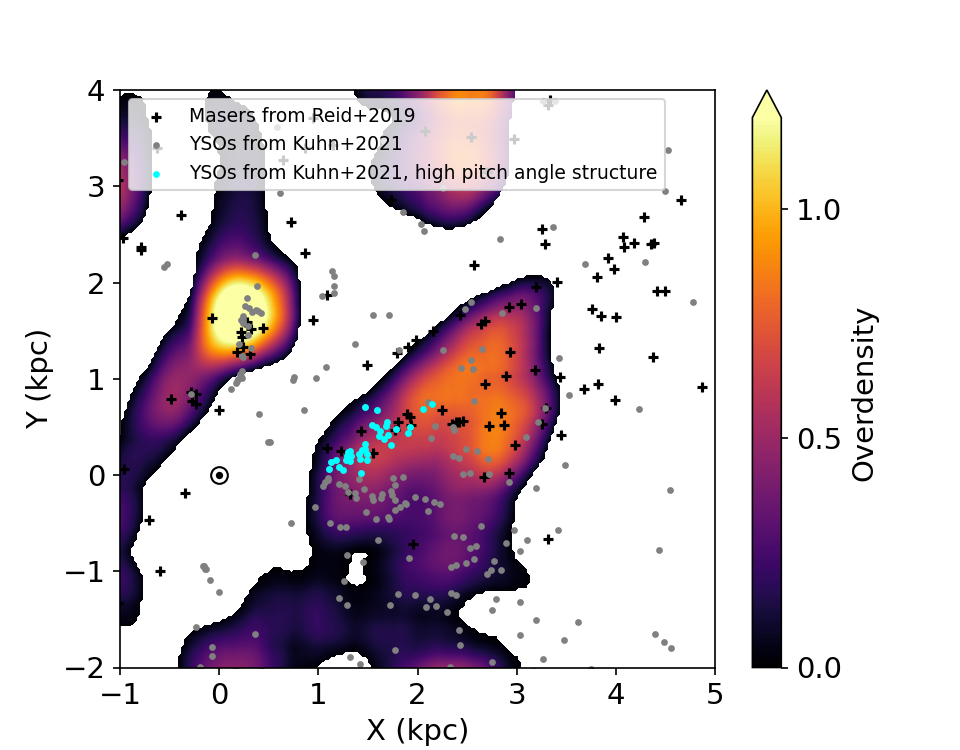}
    \caption{Comparison between the map obtained in this work (colored contours), the distribution of YSOs from \cite{Kuhn:2021_YSOcatalog} in grey (with the cyan sources identified as discussed in the text), and masers from \cite{Reid:2019}.}
    \label{fig:comparison_Kuhn}
\end{figure}

Figure \ref{fig:comparison_Kuhn} shows the comparison between our map, the distribution of Young Stellar Objects (YSOs) from \citet{Kuhn:2021_YSOcatalog} and masers from \citet{Reid:2019}, zooming in on the high-pitch angle feature discussed in detail by \citet{Kuhn:2021}. To better visualize ad discuss this region of the Galactic disc, we identify the YSOs associated with this feature by looking at the approximate location of those sources in the Galactic plane \citep[shown as yellow stars in Figure 3 of ][]{Kuhn:2021}, and highlight them in cyan in Figure \ref{fig:comparison_Kuhn} of this work. As discussed in the main text, our map shows good agreement with the observed distribution of YSOs and masers.

\bibliography{Young_Galactic_disc}{}
\bibliographystyle{aasjournalv7}

\end{document}